\documentclass[sts]{imsart}
\usepackage{caption}
\usepackage{subcaption}
\usepackage{bbm}
\usepackage{soul}
\usepackage{fix-cm}
\usepackage{anyfontsize}
\RequirePackage{amsthm,amsmath,amsfonts,amssymb}
\RequirePackage[authoryear]{natbib}
\RequirePackage[colorlinks,citecolor=blue,urlcolor=blue]{hyperref}
\RequirePackage{graphicx}
\usepackage[T1]{fontenc}
\DeclareFontShape{T1}{ptm}{m}{scit}{<->ssub*ptm/m/sc}{}

\newcommand{\cc}{{\rm C}}
\newcommand{\ct}{{\rm T}}

\usepackage{tikz}
\usetikzlibrary{positioning,chains,fit,shapes,calc}
\tikzset{fit margins/.style={/tikz/afit/.cd,#1,
    /tikz/.cd,
    inner xsep=\pgfkeysvalueof{/tikz/afit/left}+\pgfkeysvalueof{/tikz/afit/right},
    inner ysep=\pgfkeysvalueof{/tikz/afit/top}+\pgfkeysvalueof{/tikz/afit/bottom},
    xshift=-\pgfkeysvalueof{/tikz/afit/left}+\pgfkeysvalueof{/tikz/afit/right},
    yshift=-\pgfkeysvalueof{/tikz/afit/bottom}+\pgfkeysvalueof{/tikz/afit/top}},
    afit/.cd,left/.initial=2pt,right/.initial=2pt,bottom/.initial=2pt,top/.initial=2pt}

\usepackage[margin=1in]{geometry}

\usepackage{tkz-graph}
\usetikzlibrary{positioning}
\usetikzlibrary{arrows.meta}

\tikzstyle{observed}=[circle, inner sep=0mm, outer sep=0mm, minimum size=2mm, draw=black, fill=black]
\tikzstyle{unobserved}=[circle, inner sep=0mm, outer sep=0mm, minimum size=2mm, draw=black, fill=white]

\tikzstyle{notouch}=[shorten <=5pt, shorten >= 5pt, -{Latex[length=2mm, width=1.5mm]}]
\usepackage{tikz}
\usetikzlibrary{shapes,decorations,arrows,calc,arrows.meta,fit,positioning}
\tikzset{
	-latex,auto,node distance = 1.3cm and 1.3cm,
	semithick ,
	state/.style ={ellipse, draw, minimum width = 0.5 cm},
	el/.style = {inner sep=2pt, align=left, sloped},
	point/.style = {circle, draw=black, inner sep=0.04cm,
     fill=black,node contents={}},
    bidirected/.style={Latex-Latex,dashed},
}

\usepackage{comment}

\usepackage[colorinlistoftodos]{todonotes}

\newcommand{\eg}{{\it e.g., }}
\newcommand{\iiee}{{\it i.e., }}

\startlocaldefs
\theoremstyle{plain}

\theoremstyle{remark}

\newcommand{\E}{\mathbb{E}}

\newcommand{\one}{\mathbbm{1}}
\definecolor{myblue}{RGB}{80,80,160}
\definecolor{mygreen}{RGB}{80,160,80}

\endlocaldefs

\begin{document}

\begin{frontmatter}
\title{Challenges in Statistics:\\ A Dozen Challenges in Causality and Causal Inference\thanks{We are grateful to Rob Tibshirani, Anav Sood, and John Cherian for suggesting we write this paper, to Elias Bareinboim, Ivan Diaz, Vanessa Didelez, Peng Ding,  Guilherme Duarte,  Kosuke Imai, Alex Luedtke, Daniel Malinsky,  Judea Pearl, Paul Rosenbaum,  Alejandro Schuler, Dylan Small,  Liz Stuart and Stefan Wager for comments, and to the
Office of Naval Research for support under grant numbers N00014-17-1-2131 and N00014-19-1-2468, the Patient Centered Outcomes Research Initiative for award ME-2022C1-25648, and Amazon for a gift.}}
\runtitle{Challenges in Statistics: A Dozen Challenges in Causality and Causal Inference}

\begin{aug}
\author[A]{\fnms{Carlos}~\snm{Cinelli}\ead[label=e1]{cinelli@uw.edu}},
\author[B]{\fnms{Avi}~\snm{Feller}\ead[label=e2]{afeller@berkeley.edu}},
\author[C]{\fnms{Guido}~\snm{Imbens}\ead[label=e3]{imbens@stanford.edu}\orcid{0000-0002-4846-7326}},
\author[D]{\fnms{Edward}~\snm{Kennedy}\ead[label=e4]{edward@stat.cmu.edu}},
\author[E]{\fnms{Sara}~\snm{Magliacane}\ead[label=e5]{s.magliacane@uva.nl}},
\author[F]{\fnms{Jose}~\snm{Zubizarreta}\ead[label=e6]{email}}

\ead[label=u1,url]{www.foo.com}



\address[A]{Carlos Cinelli is Assistant Professor, Department of Statistics,
University of  Washington, Seattle, WA, USA\printead[presep={\ }]{e1}.}

\address[B]{Avi Feller is Professor, Goldman School of Public Policy and Department of Statistics,
University of California, Berkeley, Berkeley, CA, USA\printead[presep={\ }]{e2}.}

\address[C]{Guido Imbens is Professor, Graduate School of Business, and Department of Economics, and Director of the Standard Causal Science Center,
Stanford University, Stanford, CA, United States\printead[presep={\ }]{e3}.}

\address[D]{Edward Kennedy is Associate Professor, Department of Statistics \& Data Science,
Carnegie Mellon University, Pittsburgh, PA, United States \printead[presep={\ }]{e4}.}

\address[E]{Sara Magliacane is Assistant Professor, Informatics Institute,
University of Amsterdam, Amsterdam, Netherlands\printead[presep={\ }]{e5}.}

\address[B]{Jose Zubizarreta is Professor, Departments of Health Care Policy and Department of Biostatistics,
Harvard University, Boston, MA, USA\printead[presep={\ }]{e6}.}

\end{aug}

\begin{abstract}
Causality and causal inference have emerged as core research areas at the interface of modern statistics and domains including biomedical sciences, social sciences, computer science, and beyond. The field's inherently interdisciplinary nature---particularly the central role of incorporating domain 
knowledge---creates a rich and varied set of statistical challenges. Much progress has been made, especially in the last three decades, but there remain many open questions.
Our goal in this discussion is to outline research directions and open problems we view as particularly promising for future work. 
Throughout we emphasize that advancing causal research requires a wide range of contributions, from novel theory and methodological innovations to improved software tools and closer engagement with domain scientists and practitioners. 
\end{abstract}

\end{frontmatter}

\section{Introduction}

Causal inference is an interdisciplinary field focused on understanding cause-and-effect relationships, such as learning the effects of interventions, uncovering causal mechanisms, and providing explanations for observed phenomena. Today there is a vibrant research community studying causality in a diversity of domains---spanning biomedical sciences, the social sciences, computer science, and more---often leveraging insights from one application area to make advances in others. Statisticians have played a prominent role in causal inference, dating back to Fisher and Neyman a century ago.
As methodologists actively working in this area, our goal in this paper is to highlight both the challenges and opportunities for the next decades of statistical research in causal inference.

We are optimistic about this future, in part because the field is just getting started: despite statistics' historic role in causal inference, the broader discipline's embrace of causal inference as a core subject is relatively new.
As recently as the 1990s, there were very few scholarly articles in statistics journals that focused explicitly on causality. 
Statistics textbooks typically did not cover causal inference, beyond admonishing readers not to confuse correlation and causation, and 
any explicit mention of causality 
was typically limited to discussions of randomized experiments.\footnote{For example, in his Nobel lecture Daniel McFadden wrote,  “detection of true causal structures is beyond the reach
of statistics” and recommends that “For these reasons, it is best to avoid the language of
causality” (\citep{mcfadden2001economic}  p. 369), despite the fact that his work is very much about what we would now describe as causal.} There were no graduate or undergraduate courses devoted to causal inference,\footnote{There were courses devoted to design of experiments. As far as we know the first course devoted entirely to causal inference in experimental and observational studies was a 
graduate course Imbens and Rubin co-taught at Harvard in 1996.} 
and few presentations at major statistics conferences. There was little explicit attention to causal inference in other fields where statistical methods are used, including social and biomedical sciences. This is even more remarkable because causal effects are often precisely the quantities of interest in these fields.\footnote{Interestingly there is still pushback against this. For example,
\citet{bailey2024causal} question in general
``whether the researchers are interested in estimating a causal effect at all'' but they admit that ``in some fields the default answer may be a clear ‘yes’ (for example, in economics).''
}

\begin{figure}[!t]
\centering
\includegraphics[width=3.5in]{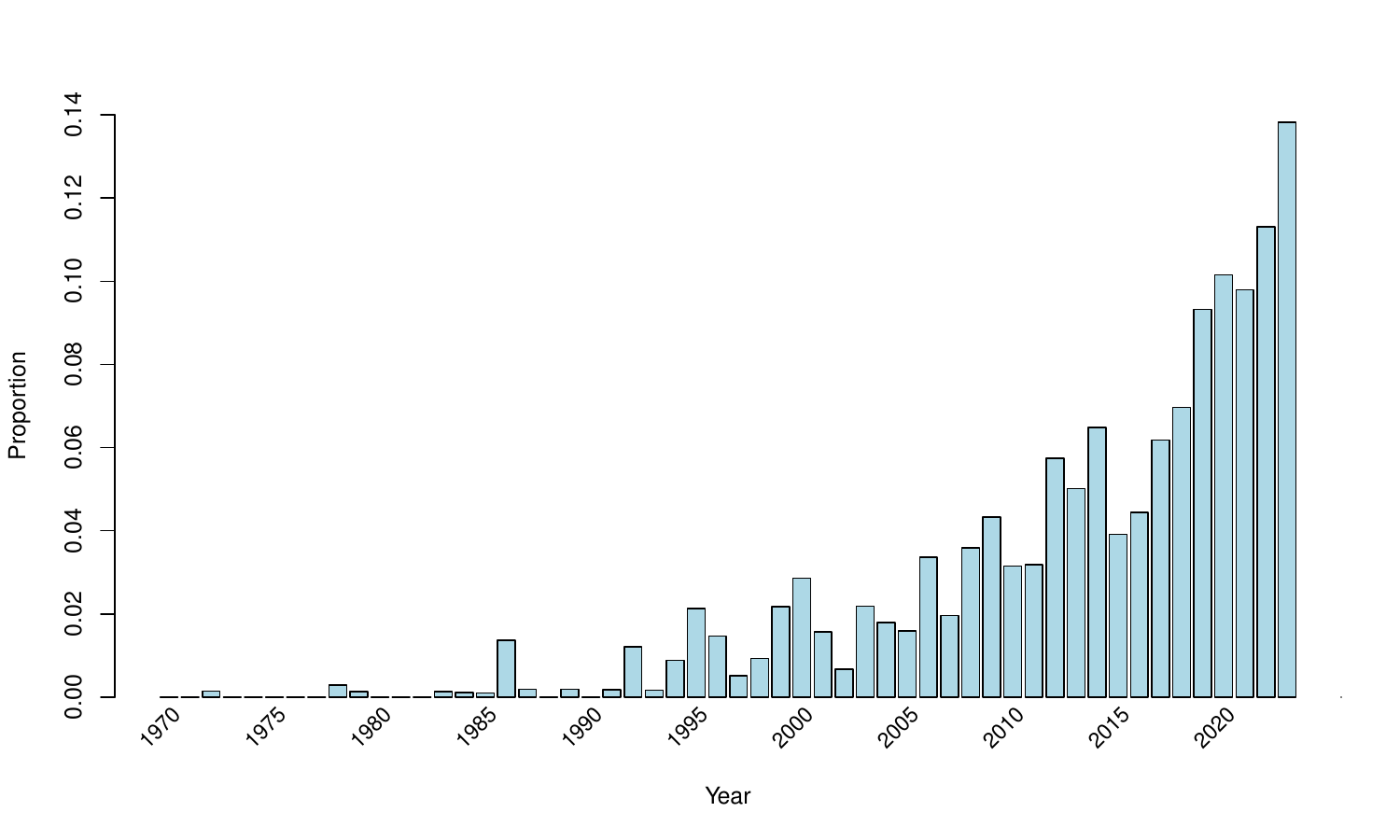}
\caption{
Proportion of articles with the word ``causal'' in the title, abstract, or keywords, in \emph{Annals of Statistics}, \emph{Annals of Applied Statistics}, \emph{Biometrics}, \emph{Biometrika}, \emph{Journal of the American Statistical Association}, \emph{Journal of the Royal Statistical Society}, and \emph{Statistical Science} from 1970 to 2024.
}
\label{fig:causal_papers}
\end{figure}

This state of affairs has changed dramatically over the last three decades.
Today many of the leading statistics journals routinely publish causal inference research (see Figure \ref{fig:causal_papers}), and 
there are now  two specialized journals devoted exclusively to the subject,
 the {\it Journal of Causal Inference} 
(since 2013) 
 and {\it Observational Studies}
 (since 2015).
Graduate and undergraduate education in statistics and substantive fields regularly includes methodological courses devoted solely to causal inference. There is a scholarly society, the {\it Society for Causal Inference}. There are multiple university centers devoted to causality and causal inference,\footnote{For example, the CAUSALab at Harvard University, the Causal Artificial Intelligence Lab at Columbia University, the Stanford Causal Science Center at Stanford University, the Center for Causal Inference at the University of Pennsylvania.} and there are annual academic conferences focused entirely on methods for and applications of causal inference.\footnote{For example, the American Causal Inference Conference (ACIC), the European Causal Inference Meeting (EuroCIM), the Causal Learning and Reasoning conference (CLeaR), and the Causal Data Science meeting.}

Given this rapid progress, the question is what challenges and open questions remain. We highlight a dozen areas below: 
\begin{enumerate}
\item Complex Experiments and  Experimental Design; 
\item Interference and Complex Systems; 
\item Heterogeneous Effects and Policy Learning;
\item Mediation and Causal Mechanisms; 
\item Optimality and Minimaxity; 
\item Sensitivity Analysis and Robustness;  
\item Reliable and Scalable Causal Discovery; 
\item Aggregation and Synthesis of Causal Knowledge; 
\item Automation of the Causal Inference Pipeline; 
\item Benchmarks, Evaluation, and Validation; 
\item New Identification Strategies; and 
\item Large Language Models and Causality.
\end{enumerate}
For each area, we provide some background and pose outstanding challenges.
We note that the challenges outlined here are not comprehensive, and, for example, \citet{mitra2022future} discuss some additional ones.

\paragraph*{What are causality and causal inference?}
Before turning to challenges and open questions, we need a working definition of causal inference and 
what separates it from other subfields of statistics; we give a more technical introduction in Section \ref{section:frameworks}. We note here that we do not draw a sharp distinction between causality and causal inference, with perhaps the term causal inference making a slightly closer link to statistical methods.
Broadly, causal inference is about answering ``what if''\footnote{The subtitle of one leading causal inference textbook, \citet{hernan2010causal}.} questions, such as learning the effect of (possibly hypothetical) interventions, uncovering causal mechanisms, or unveiling the causes of observed effects. For instance, typical examples include estimating the effect of taking a drug on some health outcome, understanding whether having a college degree was the cause of someone's earnings, or why someone had an adverse outcome. In all cases we are interested in comparing outcomes that we by definition cannot observe together---what Paul Holland in a widely quoted phrase described as ``the fundamental problem of causal inference'' \citep{holland1986statistics}. Because the focus is on comparing  (factual) outcomes that we see with (counterfactual) outcomes that we do not see, 
the challenges in causal inference differ conceptually from those in predictive modeling. Fundamentally, this comparison raises questions referred to as  \emph{identification} \citep{fisher1961identifiability}: even in large samples we may not be able to learn answers to our questions from the data at hand.\footnote{Exactly what identification means is not very well defined: \citet{leamer1983let}, in his seminal paper ``Let's take the Con out of Econometrics'' writes about a prominent researcher who ``... mentions
the phrase `identification problem,' which,
though no one knows quite what he means,
is said with such authority that it is totally
convincing.''} For instance, simply comparing observed outcomes for units exposed to different levels of a cause may tell us little about the causal effect due to \emph{confounding}---units may differ across levels of the cause in ways related to the outcome. Moreover, these confounders are not always observed by the researcher, raising severe challenges for learning about the causal effects. 

Dealing with these challenges requires assumptions that are substantive, or more precisely, causal---as opposed to solely statistical---in nature. Such identifying assumptions are by their nature untestable and require strong domain knowledge for drawing compelling, credible, causal conclusions. This makes causal inference intrinsically interdisciplinary, with models and methods developed by researchers in a variety of fields outside statistics for their specific substantive discipline, including computer science, political science, biomedical science, economics, and law.
The interdisciplinary nature of causality and causal inference creates both challenges and opportunities. 
Differences in methods and terminology between fields can  make it harder to recognize the commonalities in questions and problems. At the same time, insights from one discipline may have applications in other disciplines.

\paragraph*{Where has causal inference been successful?}
Arguably, the most impactful research in causal inference has been on the design and analysis of randomized experiments, going back to the 1920s and 1930s. In particular, 
\citet{fisher1937design} and \citet{neyman1923} showed the power of randomization in inferring causal effects, initially in the context of agricultural experiments. These ideas ultimately led the Food and Drug Administration (FDA) in the United States to insist on randomized experiments as part of the drug approval process  \citep{hill1990memories}. Today randomization remains the gold standard for causal inference. It has moved from being largely confined to biomedical
settings to now being a staple of data analyses much more broadly, especially online in the private  sector. For instance, large tech  companies conduct hundreds of thousands of experiments every year to improve their  products and services \citep{gupta2019top}.

More broadly, causal inference has had repeated success in sharpening substantive research questions, often by clarifying the underlying assumptions and the target quantity of interest. This has proved especially useful in applying novel methods from one field in others. 
A leading example
is the \emph{regression discontinuity design} (RDD), which was first developed by education researchers \citep{thistlethwaite1960regression} and which remained a niche topic for decades \citep{cook2008waiting}.
Today, RDD is used widely across the social and biomedical sciences, largely driven by a common causal inference framework that allows researchers in otherwise disparate domains to understand and apply the same methods.
Another prominent example is \emph{instrumental variables} methods, which originated in economics \citep{wright1928tariff, tinbergen1930determination}\footnote{Scholars debate whether the economist Philip Wright or his son, the geneticist Sewall Wright and inventor of path analysis, wrote the first exposition of instrumental variables. See \cite{stock2003retrospectives} for stylometric evidence suggesting it was indeed Philip Wright.}, and spread throughout statistics in the 1990s \citep{angrist1996identification} and subsequently found new applications in biomedical settings through Mendelian randomization \citep{sanderson2022mendelian}.

Finally, there are also many celebrated analyses that use the tools of causal inference. 
An early study in epidemiology, \citet{snow1856mode} inferred the causal mechanism for the spread of cholera through carefully documenting the location of cholera cases in relation to the pumps used for obtaining drinking water, in what would now be called a natural experiment.\footnote{See \citet{Angrist:2010} for a general discussion of natural experiments and the so-called credibility revolution in econometrics.} Another classic example is assessing the causal effect of smoking on cancer, where a variety of studies documenting different pieces of evidence ultimately led to the consensus that there was indeed a causal relationship \citep{cornfield1959smoking}, despite the presence of detractors among leading statisticians, including Fisher \citep{fisher1958cigarettes}.

\paragraph*{Where has causal inference failed?}
There have also been many failures of studies of causal inference. 
Often researchers have extrapolated insights from a study population to a target population that is quite different. For example, thalidomide had been prescribed as a sleep aid, but once it was prescribed to pregnant women it was found to lead to birth defects \citep{melchert2007thalidomide}. These failures of {\it external validity} can plague both experimental and observational studies.

Another type of failure arises when researchers place excessive confidence in the assumptions required to identify causal effects. 
A prevalent example is assuming that all relevant confounders are measured without sufficient justification or adequate sensitivity analyses.
Ed Leamer argued that ``Hardly anyone takes data analyses seriously''  \citet[p. 37]{leamer1983let}, emphasizing the limited credibility of much empirical work in the social sciences, especially around the question of the deterrence effect of the death penalty on murder rates. 
The assumptions and models used to justify such estimates are often strong and arguably implausible; nonetheless, the full extent of this issue can go unnoticed.\footnote{This body of work is sometimes referred to as the ``credibility crisis in economics'' \citep{Angrist:2010}, a precursor to what Angrist and Pischke later termed the ``credibility revolution.'' It is also related to the broader ``replication crisis'' in the social sciences, although that crisis raises other issues besides causality.}

Finally, causal inference has frequently been used ``for support rather than illumination,'' regulating science instead of supporting it \citep[see, for example,][]{recht2025bureaucratic}. One consequence is that whole groups of researchers and modes of inquiry ({\it e.g.}, qualitative methods and case studies) have often been excluded from the field, leaving the discipline poorer and limiting the potential reach of causal inference ideas.

\paragraph*{Organization.}
The rest of the paper is organized as follows. In Section \ref{section:frameworks} we describe  current perspectives  on causal inference and the state of  the literature.
In Section \ref{section:challenges} we describe some of the main outstanding challenges for causal inference.
In Section \ref{section:thoughts} we provide some concluding thoughts.

\section{The Current State of Causal Inference}
\label{section:frameworks}

To set the stage for the outstanding challenges and future directions for research on causal inference we briefly introduce the standard frameworks for causal inference, and then describe the current state of the field. More detailed discussions of these issues can be found in a quickly increasing number of textbooks on causal inference, including \cite{cunningham2018causal, ding2024first, hernan2010causal, 
huntington2021effect, imbens2015causal, morgan2015counterfactuals,pearl2009causality, pearl2018book,rosenbaum2002observational, rosenbaum2020design} and  \citet{wager2024}.

\subsection{Frameworks for Defining Causal Effects}

One of the key achievements of modern causal inference research was the development of formal frameworks for articulating causal questions and causal assumptions, as well as deriving causal answers. Though their usage may differ to reflect traditions and idiosyncrasies of various fields, these frameworks  are now broadly taught, acting as a common language for scientists across  disciplines for understanding and discussing causal inference problems. Here we briefly introduce three closely related  approaches, or languages, for causal inference used in the literature, highlighting some of their common aspects.

\paragraph*{Potential Outcomes.} Dating back to the 1920s \citep{neyman1923}, the \emph{potential outcomes framework} originated in studies of randomized experiments. Donald Rubin subsequently generalized this framework to aid in the design and analysis of observational studies \citep{Rubin:1974}; see \citet{Imbens:Rubin:2015} for a textbook based on this formulation. 

This framework starts with the notion that each unit has multiple potential outcomes, each corresponding to a level of the treatment; these potential outcomes are the key objects in this approach. With a binary treatment $W_i\in\{0,1\}$ and with some additional restrictions (see below), the two potential outcomes are $Y_i(0)$ and $Y_i(1)$, for the outcomes without and with treatment respectively. Because the treatment can only take on one value for a particular unit (\eg for a specific individual at a specific time), at best only one of these potential outcomes can be observed, leading to Holland's celebrated ``fundamental problem of causal inference.'' The realized and possibly observed outcome is the potential outcome corresponding to the treatment received, $Y_i\equiv Y_i(W_i)$. Causal effects are then defined as comparisons of these potential outcomes, \eg$Y_i(1)-Y_i(0)$. Assumptions are often formulated in terms of  an {\it unconfoundedness} or \emph{ignorability}  assumption, \citep{rosenbaum1983central}:
\begin{align}
 W_i \perp\!\!\!\!\perp \Bigl(Y_i(1), Y_i(0)\Bigr) \  \Bigl|\ X_i. \label{eq:ignorability}    
\end{align}

These assumptions entail conditional independence restrictions on the relationship between the level of the treatment and the potential outcomes, given some set of observed covariates or pretreatment variables $X_i$. For example, in a randomized experiment, the treatment would be guaranteed to be independent of potential outcomes by design, and thus the treatment assignment is ignorable or unconfounded. Other assumptions commonly used in the potential outcome framework include ruling out particular causal effects, so-called \emph{exclusion restrictions}, or shape restrictions such as {\it monotonicity.}
For example, in an instrumental variables setting 
\citep{imbens1994,angrist1996identification}, one might start with potential outcomes $Y_{i}(w,z)$ indexed by the treatment $W_i$ and an instrument $Z_i$, and make the assumption that the potential outcomes do not depend on the instrument, or $Y_i(w,z)=Y_i(w,z')$ for all $z,z'$. In addition, in the instrumental variables setting one postulates potential outcomes for the treatment, $W_i(z)$, for $z=0,1$, with the monotonicity assumption that $W_i(1)\geq W_i(0).$

\paragraph*{Structural Equation Models.}
Another framework for causal inference is to model mechanisms via a system of \emph{structural equations}. 
This approach originated in the econometrics literature \citep{haavelmo1943statistical,strotz1960recursive}, with further development in the social sciences \citep{bollen1989structural,duncan2014introduction} and in computer science \citep{pearl2009causality}. Unlike the potential outcomes framework, this approach models the data generating process through a series of independent and local mechanisms, called structural equations. Interventions and causal effects in this framework are defined as the result of modifications on (parts of the) structural equations, while keeping the rest of the system untouched. While models were traditionally assumed to be linear, these can be fully nonparametric in modern treatments. Importantly, nonparametric structural equation models can be shown to be formally equivalent to the potential outcomes framework under some common assumptions \citep{pearl2009causality}. 

For instance, paralleling the previous example with unconfoundedness as in Equation (\ref{eq:ignorability}),  consider the following nonparametric structural equation model: 
\begin{align}
    X_i &= f_x(U_{xi}) \nonumber\\
    W_i &=  f_w(X_i, U_{wi}) \nonumber\\
    Y_i &=  f_y(W_i, X_i, U_{yi}). \nonumber
\end{align}
Here $U_x$, $U_w$ and $U_y$ stand for unobserved factors not modeled by the analyst, independent of each other. In this model, the potential outcome $Y_i(w)$ is defined as the solution of the system in which $W_i$ is set experimentally to $w$, {\it i.e.}, $Y_i(w)\equiv f_y(w, X_i, U_{yi})$. The assumption that unobserved variables are mutually independent implies the ignorability assumption (\ref{eq:ignorability}). Exclusion restrictions in structural models are represented by the absence of a variable in the structural equation of another. Shape restrictions, such as linearity or monotonicity, can be  encoded as constraints on the form of structural equations.

\paragraph*{Graphical models.}
Graphical models or Directed Acyclical Graphs (DAGs) have emerged as a third language to encode causal assumptions, as well as to derive their logical ramifications, in an intuitive  manner.  Originating in the work of Sewall Wright \citep{wright1934method}, with additional discussions in economics, \eg\citet{tinbergen1940econometric, griliches1972education}, 
their modern form is largely due to Peter Spirtes, Clark Glymour \citep{Glymour1987-GLYDCS,spirtes2001causation}, and Judea Pearl \citep{pearl1995causal, pearl2009causality}. In a causal graph, an arrow $W_i \rightarrow Y_i$ denotes that $W_i$ may be a direct cause of $Y_i$. Its most general formulation does not impose any functional form or distributional assumptions about this causal relationship. 

\begin{figure} 
    \begin{tikzpicture}[
        >=stealth,
        node distance=2.0cm
        ]
        \node[observed, label=below:{\(W\)}] (1) at (0,0) {};
        \node[observed, right=of 1,  label=below:{\(Y\)}] (2) {};
        \node[observed, left=of 1, label=below:{\(Z\)}] (4) {};
        \node[unobserved, label=above:{\(U\)}] (3) at ($(1)!0.5!(2)+(0,1)$) {};
        
        \draw [->, notouch] (1.east) -- (2.west);
        \draw [dashed, ->, notouch] (3.south west) -- (1.north east);
        \draw [dashed, ->, notouch] (3.south east) -- (2.north west);
        \draw [->, notouch] (4.east) -- node[above] {$+$} (1.west);
    \end{tikzpicture}
\caption{Example of a causal diagram for an instrumental variable model.}
\label{fig:iv}
\end{figure}
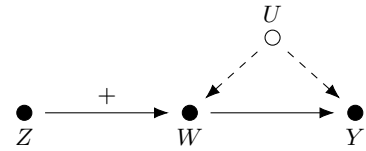

Figure~\ref{fig:iv} shows a typical causal diagram for an instrumental variables setting. The focus here is on the causal effect of the treatment $W$ on the outcome $Y$. The complication is that there is an \emph{unobserved confounder} $U$ that affects both the treatment and the outcome, invalidating a direct comparison of units by treatment status. The instrument $Z$, which affects the treatment but has no direct effect on the outcome, can help in the identification of causal effects of the treatment on the outcome.  The exclusion restriction is encoded by the absence of a direct path or  arrow between $Z$ and $Y$ (although there is an indirect path from $Z$ to $Y$  through $W$); unconfoundedness of the instrument is encoded by the absence of arrows between $U$ and $Z$ and no unmeasured confounders of the instrument. Until recently, it has been difficult to graphically encode the monotonicity assumption, {\it  i.e.}, that the effect of the instrument $Z$ on the treatment $W$ is non-negative for all units; there are now several proposals \citep{vanderweele2010signed,maiticounterfactual}, including by annotating arrows with a `+' sign. Finally, potential outcomes can also be directly represented in causal graphs; see for example, Twin Networks \citep{pearl2009causality}, Single World Intervention Graphs (SWIGs) \citep{richardson2013single}, or Ancestral Multi-World Networks (AMWN) \citep{correa2025counterfactual}. 

One way to connect the graphical model of Figure~\ref{fig:iv} with the previously discussed structural equations framework is to read it as encoding the following system of nonparametric structural equations,
\begin{align}
    Z_i &= f_z(U_{zi}) \nonumber\\
    W_i &=  f_w(Z_i, U_i) \nonumber\\
    Y_i &=  f_y(W_i, U_i), \nonumber
\end{align}
along with the assumption that $U_z$ and $U$ are independent. Translating to potential outcomes, note this model encodes both the exclusion restriction, {\it  i.e.}, $Y_i(w, z)\equiv  f_y(w, U_i)=Y_i(w)$, as well unconfoundedness of the instrument, \iiee $Y_i(w,z), W_i(z) \perp\!\!\!\!\perp Z_i$. The monotonicity assumption, denoted in the graph by $Z_i \xrightarrow{+} W_i$, can be interpreted as the functional constraint in the structural equations form, $f_w(1, u) > f_w(0, u),\forall~ u$, or, in the potential outcomes form, $W_i(1)>W_i(0), \forall~i$. 

Overall, similarly to how we can use different languages to express the same reality, these different languages for causal inference  allow us to express the data generating process in various ways, and each of them can provide a more precise or convenient terminology depending on the setting. Understanding the connections between these frameworks, as well as possibly developing new frameworks and building blocks of causality will continue to be critical for advancing the field.

\subsection{The central historical role of the case of binary treatments under unconfoundedness}
\label{section:state}

Much of the causal inference literature from the 1980s onward has focused on a relatively narrow problem: how to estimate the average causal effect of a binary treatment on a scalar outcome in a setting where assignment to the treatment is ignorable; {\it i.e.}, a setting with no unobserved confounders.  A key paper is \citet{rosenbaum1983central}, which introduced the ignorability assumption and the {\it propensity score}, the conditional probability of being exposed to the treatment given the confounders.\footnote{The Rosenbaum-Rubin paper has now, in 2025, almost 40,000 google scholar cites, accumulating currently  at a still increasing rate of 3,000 per year.}

This setting has led both to a vast and successful theoretical literature and to a huge  empirical literature. Researchers have proposed a remarkably large number of different estimation approaches, including
outcome modeling, inverse propensity score weighting, matching, and doubly robust methods based on combinations of outcome modeling and inverse propensity score weighting. Many of these are widely used in empirical practice, although very simple methods based on least squares analyses continue to be the most popular. See any of the modern textbooks for detailed discussions of these methods.

The corresponding theoretical literature has been very successful partly because it focused on a common problem. The drawback has been that it has also remained narrow. A key theme in our discussion is not only in pushing more causal inference research beyond this setting, but also in speeding 
up the practical adoption of existing causal inference advances beyond this binary case.

There are three ways in which this focus has been narrow.
First, most of the work focused on the case with all confounders observed. This is unlikely the case in practice; we discuss many alternatives below.
Second, most of the work has focused on the setting with a single binary cause. In practice, causes are often much richer, involving multiple causes, causes taking on many values, and complex systems.
Third, the literature has also largely limited itself to estimating the effects of causes: what is the effect of a particular intervention. A different set of questions focuses on than causes of effects \citep{Gelman:Imbens:2013} or questions about attribution \citep{yamamoto2012understanding}: why did this happen, or how much of what we see can we attribute to different causes. In practice there is also widespread interest in such questions.

\section{Causal Inference: A look to the future}
\label{section:challenges}

In this section we discuss a dozen areas that we view as promising for future research on causality. In each case we lay out the general setting, and then discuss some specific challenges.

\subsection{Complex Experiments and Experimental Design}
\label{section:experiments}

\subsubsection{Motivation}
For most of the 20th century, the causal inference literature on experimental design centered on agricultural and biomedical settings, typically with a modest number of fixed units ({\it e.g.}, patients or plots of land) and a small number of treatments (\eg active drug versus placebo); see \citet{wu2011experiments} for a modern discussion.\footnote{The literature on design of (often factorial) experiments in industrial settings was largely distinct.}
Over the past twenty years, however, the practice of experimentation has changed dramatically from these classical origins, driven in part by new technologies and the rapid growth of the tech sector. 
Today, many experiments take place in online environments and are run by private organizations \citep{gupta2019top}---a sharp departure from, among others, clinical trials that require FDA regulatory approval.
Experimentation in offline environments can also be much more complex, such as \emph{when} to treat in mobile health interventions \citep{nahum2018just}. 
These modern experimentation settings raise new challenges and fundamental questions, and exhibit deeper connections with the parallel literature in machine learning.

\subsubsection{Background.}
\paragraph*{Online and adaptive experimentation.}  
In many modern settings units enter into the experiment sequentially, and outcomes can be measured soon after exposure. This combination of sequential enrollment and fast observation greatly expands the design space, especially by enabling researchers to adapt the design during the experiment. While these ideas have a long history in statistics \citep{efron1971forcing} and in the bandit literature \citep{lai1985asymptotically}, in recent years there has been a large and very active literature on multi-armed bandits and sequential experimentation \citep{lattimore2020bandit}. For example, one prominent research direction is allowing the experimental design and analysis to vary across individual-level covariates, often referred to as ``context'' in this literature 
\citep{agarwal2014taming, dimakopoulou2019balanced}.
Another recent push has focused on estimating average treatment effects after adaptive experiments \citep[\eg][]{hahn2011adaptive, kato2020efficient}. Finally, there have been multiple extensions to settings with complex causal models, both with known
\citep{lee2018structural} and unknown causal structure \citep{lu2021causal, bilodeau2022adaptively}.

\paragraph*{Complex experimental designs.} A related challenge is that researchers today can control many more aspects of complex experiments. For instance, many modern experiments have a large number of possibly interacting treatments \citep{zhao2022regression}, such as in conjoint experiments popular in survey research \citep{bansak2021conjoint}. This added complexity is also more common in non-randomized studies, in which greater care is needed to investigate interacting treatments \citep{yu2023balancing}.
Increasingly, researchers also have control over the timing of treatment, such as in mobile health interventions \citep{nahum2018just, kirgios2025does} or when learning treatment policies \citep{nie2021learning}.
Finally, the treatments themselves are becoming much more complex, especially with the increasing use of text and images as treatments \citep{egami2022make}; see Section \ref{sec:llm}.

\subsubsection{Challenges}
\paragraph*{Sequential treatments and increased adaptivity.}
The adaptive experimentation literature has largely focused on the case with a single treatment per unit. Current adaptive methods are largely confined to settings with few constraints on the adaptivity. 
Much more challenging are settings with sequential treatments. The initial work in this area has relied heavily on detailed knowledge of the causal structure. In practice combining these questions with more limited knowledge about the causal structure will be important.
Moreover, the adaptive nature of the experiments also raises questions about appropriately adapting always valid confidence sequences that are popular in online experimentation. Finally, there remain under-explored links to reinforcement learning \citep{sutton1998reinforcement}.

\paragraph*{Platform trials.}
In connection with adaptive experiments, platform trials have gained considerable traction in the health sciences as a means to study new medical interventions in a timely manner.
Unlike traditional trials that focus on specific treatments, platform trials center around particular diseases, allowing the evaluation of multiple interventions as they become available.
These trials operate under a master protocol and often involve multiple sites, which helps in collecting broader samples and enhancing generalizability.
Platform trials present interesting methodological challenges including the definition of estimands, robust estimation, and multiple comparisons \citep{santacatterina2025identification, qian2024estimands}.

\paragraph*{Automation and reducing frictions.}
In practice many frictions prevent researchers from implementing adaptive experimentation. Even if only some of these barriers are ``statistical'' in a traditional sense, our field is still responsible for working to reduce these barriers. 
One promising direction is the increased use of Large Language Models and AI agents to automate more complex designs; see Section \ref{sec:llm}. This introduces many additional challenges, including how to adapt pre-analysis plans and other best practices beyond simple settings.

\subsection{Interference and Complex Systems}

\subsubsection{Motivation}

The classical causal inference paradigm assumes \emph{no interference} between units \citep{cox1958planning}: a unit's outcome only depends on that unit's own treatment. This is typically formalized via the unfortunately-named Stable Unit Treatment Value Assumption \citep[SUTVA;][]{rubin1980randomization}, that is, the potential outcomes depend only on the assignment for unit $i$: $Y_i(\mathbf{W}) = Y_i(W_i)$, where $\mathbf{W} = (W_i, \mathbf{W}_{-i})$ is the vector of treatment assignments for all units and $\mathbf{W}_{-i}$ is the corresponding vector excluding unit $i$.
Historically, spillovers and other violations of the no-interference assumption were regarded as nuisances and were relatively rare. 
In agricultural experiments, plots of land do not interact with each other, at least after introducing sufficient space to avoid spillovers between adjacent plots; and
in most classical medical settings, interventions applied to one patient do not affect other patients. 
Interference is a major concern in vaccine trials for infectious diseases: epidemiological concepts like \emph{herd immunity} inherently violate the no-interference assumption. 
In such settings, interference need not be a nuisance but may be itself the quantity of interest.
 
In social science settings, however, spillovers and interference are often the rule, rather than the exception, putting the social into social sciences, and they are often the primary objects of interest.
For example, tutoring some students in a class may affect their classmates not enrolled in the program.
Providing information about new technologies to some farmers may lead them to share the information with other farmers not directly exposed to the new information.  
Changing the information about some rental properties in an online property rental environment
may shift demand from control to treated properties, affecting the market equilibrium and so indirectly changing the experience for control properties.

\subsubsection{Background}
Causal inference under interference is typically impossible to solve without additional structure: in a single experiment we observe only outcomes for a single vector of treatment assignments.  
Fortunately, there has been substantial progress on the vast area of restricted interference that lies between the extremes of no interference and arbitrarily complex interference, building off pioneering work from \citet{sobel2006randomized}, \citet{hudgens2008toward} and \citet{manski2013identification}. 
One increasingly common technical tool is to define \emph{exposure mappings} \citep{aronowsamii, manski2013identification}, low-dimensional summaries of the full treatment assignment vector $\mathbf{W}$ and unit characteristics that capture how each unit is affected by the treatment status of other units. These summaries are then tailored to specific problems.\footnote{
A complementary research thread instead asks what estimation is possible under potentially arbitrary interference or spillovers, viewing interference as a nuisance rather than a quantity of interest; see, for example, \citet{savje2021average} and \citet{leung2022causal}.}
For instance, in a social network setting, interference could be well described by the average characteristics of an individual's friends \citep{ogburn2024causal}.
In a marketplace, interference might be characterized by a summary measure like price \citep{munro2021treatment}. In an education setting, spillovers can be summarized via the fraction of treated units in a peer group \citep{manski1993identification, carrell2013natural}. 
Finding effective structures for interactions and spillovers---and incorporating those structures or frameworks into other aspects of causal inference, such as experimental design---is an active area of research. Importantly, these questions typically
involve delicate subject matter knowledge and are inherently interdisciplinary.

\paragraph*{Example: Bipartite Graph.} 
Consider the \emph{bipartite graph} setting 
with two populations of units as illustrated in Figure \ref{fig:full_bipartite_graph}; one prominent example is \citet{papadogeorgou2019causal}, which considers the impact of reducing power plant emissions on ambient ozone. 
The researcher can intervene on the first set of unit, the so-called \emph{intervention units} (\eg\ power plants) by exposing them to different levels of treatment. Outcomes, however, are observed for a second set of units, the \emph{outcome units} (\eg sensor locations).
Intervention units may affect partially overlapping subsets of the outcome units. 
Designing experiments that allow researchers to learn effectively about policies of interest that have this form remains an open question \citep{lu2025design}.

{\small
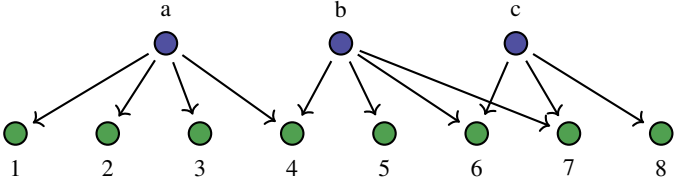
\begin{figure}[!t]
\centering

\begin{tikzpicture}[thick,
  every node/.style={draw,circle},
  fsnode/.style={fill=myblue},
  ssnode/.style={fill=mygreen},
  every fit/.style={rectangle,draw},
  -,shorten >= 3pt,shorten <= 3pt
]

\begin{scope}[start chain=going right,node distance=20mm]
\foreach \i in {a,b,c}
  \node[fsnode,on chain] (f\i) [label=above: \i] {};
\end{scope}

\begin{scope}[xshift=-2cm,yshift=-1.2cm,start chain=going right,node distance=9mm]
\foreach \i in {1,2,...,8}
  \node[ssnode,on chain] (s\i) [label=below: \i] {};
\end{scope}

\foreach \i in {1,2,3,4} \draw[->] (fa) edge (s\i);
\foreach \i in {4,5,6,7} \draw[->] (fb) edge (s\i);
\foreach \i in {6,7,8} \draw[->] (fc) edge (s\i);
\end{tikzpicture}

\caption{General bipartite interference graph with 3 interventional units (labeled a, b, c) and 8 outcome units (labeled 1, 2, 3, 4, 5, 6, 7, 8).}
\label{fig:full_bipartite_graph}

\end{figure}

}

\paragraph*{Example: Experiments in Two-sided Marketplaces.}
Consider a marketplace where the researcher is interested in estimating the average effect of providing additional information  to customers purchasing services or items. For example, Airbnb may provide videos or reviews to potential renters for some properties/providers. 
Marketplaces such as Airbnb  in which many customers interact with many properties, have complex spillovers and are poorly suited to traditional experiments. Consider a traditional experiment that assigns some properties to the treatment group and some to the control group.
Such experiments may lead customers to switch from control properties to treated properties without otherwise change a customer's overall rental level. Thus, simply reporting a difference in average outcomes could lead to misleading results with regard to the overall average effect of switching all properties between control and treatment.
\citet{bajari2023experimental, johari2022experimental} instead consider a more complex randomization that assigns pairs of units---pairs consisting of one property  and one customer---to treatment or control, as in the assignment matrix in Figure \ref{fig:market_experiment_assignment}.
Here some properties are assigned to a customer experiment where---for those properties only---all customers are randomly assigned to treatment or control. The remaining properties are assigned to a property experiment where the properties are randomly assigned to treatment or control.
This randomization creates systematic variation in the fraction of treated properties for each customer and in the fraction of treated customers for each property, both of which enable direct estimation of the causal quantities of interest.

\begin{figure}[htb]
{\small
\[
\left(
\begin{array}{llcccccccccccccccc}
&& \multicolumn{8}{c}{\rm Customer} &&\multicolumn{6}{c}{\rm Property} \\
&& \multicolumn{8}{c}{\rm Experiment} &&\multicolumn{6}{c}{\rm Experiment} \\
{\rm Properties} \rightarrow& &1 & 2 & 3 & 4 & 5 & 6 & 7 & 8 & & 9  & 10 & 11 & 12& 13 & 14\\
{\rm Customers} & 
\\
\downarrow & \\
1&   &\textcolor{green}{\cc} & \textcolor{green}{\cc} &\textcolor{green}{\cc} & \textcolor{green}{\cc} &\textcolor{green}{\cc} & \textcolor{green}{\cc}  &\textcolor{green}{\cc}  & \textcolor{green}{\cc}& & \ct& \textcolor{red}{\cc} &\ct&\textcolor{red}{\cc} & \textcolor{red}{\cc} &\ct\\
2&& \textcolor{green}{\cc} & \textcolor{green}{\cc} &\textcolor{green}{\cc} & \textcolor{green}{\cc} &\textcolor{green}{\cc} & \textcolor{green}{\cc} & \textcolor{green}{\cc}  & \textcolor{green}{\cc}& & \ct& \textcolor{red}{\cc} &\ct&\textcolor{red}{\cc}& \textcolor{red}{\cc} &\ct \\
3 & & \ct  & \ct & \ct  & \ct &\ct  & \ct & \ct   &\ct && \ct& \textcolor{blue}{\cc} &\ct&\textcolor{blue}{\cc}& \textcolor{blue}{\cc} &\ct \\
4 &&\textcolor{green}{\cc} & \textcolor{green}{\cc} &\textcolor{green}{\cc} & \textcolor{green}{\cc} &\textcolor{green}{\cc} & \textcolor{green}{\cc} & \textcolor{green}{\cc}  & \textcolor{green}{\cc}& & \ct&\textcolor{red}{\cc} &\ct&\textcolor{red}{\cc}& \textcolor{red}{\cc} &\ct\\
5&& \ct  & \ct &\ct  & \ct &\ct  & \ct & \ct   & \ct &&\ct& \textcolor{blue}{\cc} &\ct&\textcolor{blue}{\cc}& \textcolor{blue}{\cc} &\ct\\
6& & \ct  & \ct &\ct  & \ct &\ct  & \ct & \ct   & \ct && \ct&\textcolor{blue}{\cc} &\ct&\textcolor{blue}{\cc}& \textcolor{blue}{\cc} &\ct\\
7&  &\textcolor{green}{\cc} & \textcolor{green}{\cc} &\textcolor{green}{\cc} & \textcolor{green}{\cc} &\textcolor{green}{\cc} & \textcolor{green}{\cc} & \textcolor{green}{\cc}  & \textcolor{green}{\cc} && \ct&\textcolor{red}{\cc} &\ct&\textcolor{red}{\cc}& \textcolor{red}{\cc} &\ct\\
8 & & \ct  & \ct &\ct  & \ct &\ct  & \ct &   \ct & \ct && \ct&\textcolor{blue}{\cc} &\ct&\textcolor{blue}{\cc}& \textcolor{blue}{\cc} &\ct\\
9&  & \ct  & \ct &\ct  & \ct &\ct  & \ct & \ct   & \ct && \ct&\textcolor{blue}{\cc} &\ct&\textcolor{blue}{\cc}& \textcolor{blue}{\cc} &\ct\\
10& & \textcolor{green}{\cc} & \textcolor{green}{\cc} &\textcolor{green}{\cc} & \textcolor{green}{\cc} &\textcolor{green}{\cc} & \textcolor{green}{\cc} & \textcolor{green}{\cc}  & \textcolor{green}{\cc} && \ct&\textcolor{red}{\cc} &\ct&\textcolor{red}{\cc}& \textcolor{red}{\cc} &\ct\\
\end{array}
\right)
\]
}
\caption{Possible randomization for customer and properties in a marketplace, following \citet{bajari2023experimental}. \label{fig:market_experiment_assignment}}
\end{figure}

\subsubsection{Challenges.}
Causal inference with interference is fundamentally an \emph{Anna Karenina} problem, as in the opening line of the Tolstoy classic: ``All happy families are alike; each unhappy family is unhappy in its own way.'' Analogously, all no-interference applications are alike; each SUTVA violation violates SUTVA in its own way. 

\paragraph*{Balancing generality and practicality.} 
One of the greatest challenge in this research area is to develop approaches that are general enough to apply to the many different types of interference applications, while also tackling important practical complications with each distinct type. Despite rapid progress in the past decade, the currently fragmented nature of the area has made it difficult for practitioners to see what is relevant for empirical work and to incorporate the latest advances into their research.
For example, one divide is between interference  that only depends on other units' treatment assignment, which is often the focus  in the statistics literature \citep{hudgens2008toward}, and interference that depends on other units' treatments \emph{and} outcomes, which is more common in the economics literature on peer effects \citep{manski1993, sacerdote}. 
Another divide is between research based primarily on potential outcomes rather than the extensive causal discovery and graphical causal model literature on interference, which primarily focuses on identification \citep[see, for example][]{sherman2018identification, sherman2020dynamic_interference}. While there has been recent progress bridging these areas, we expect substantial gains from further work.

\paragraph*{Incorporating substantive theory.}
Another central challenge is to better incorporate substantive theory into the analyses. 
For example, spillovers that affect units far away in some metric are rare but often of great importance---and finding such cascades has proved difficult empirically. Incorporating microfoundation network models of cascades could improve the ability to detect such spillovers. One issue is that the theory often requires subject matter knowledge beyond the statistical expertise required for causal inference in settings without interference.
A related concern is the difficulty of measuring spillovers, and reflecting any measurement difficulties in the resulting analysis \citep{bhattacharya2020causal}.

\paragraph*{Interference in observational studies.} A third set of challenges concerns generalizing the methods that have been developed for experimental settings to the observational ones that are much more common. 
Understanding the experimental case with known assignment mechanisms is just the first step: spillovers are common in observational settings where challenges are bigger.
For instance, in the standard no-interference case, addressing confounding typically involves adjusting for unit-specific characteristics. In the setting with spillovers the characteristics of other units may also matter, possibly making standard adjustment procedures insufficient.

\paragraph*{Time evolution.}

The nature of interference may change over time. For example in \citet{christakis}, the authors have data on network connections at multiple points in time and find that the network linking individuals changes substantially. This may be in response to treatments, or in response to outcomes, in both cases requiring subtle adjustments to estimators for causal effects. See \citet{gao2024endogenous} for a discussion in the context of randomized experiments.

\paragraph*{Statistical power.} 
Statistical power remains a core issue with all but the largest empirical analyses of interference. A key challenge is to develop tools for applied researchers to better assess power in practice and to think about the appropriate notion of ``replicates'' under interference: intuitively, there is limited power to detect an interference pattern of interest that only has a few replicates.
This question of replicates is closely related to the issue of external validity of causal inference studies with interference. While this is an issue in all studies, this is especially challenging without the standard formulation of \textit{iid} draws from a common population.
One interesting direction is to use adaptive designs based on the propensity score \citep{hahn2011adaptive}, or, in network settings, to use graph-cutting techniques to balance bias and variance
\citep{taylor2018randomized}.

\subsection{Heterogeneous Effects and Policy Learning}

\subsubsection{Motivation}

\begin{figure}[!ht] 
\centering
\includegraphics[width=3.4in]{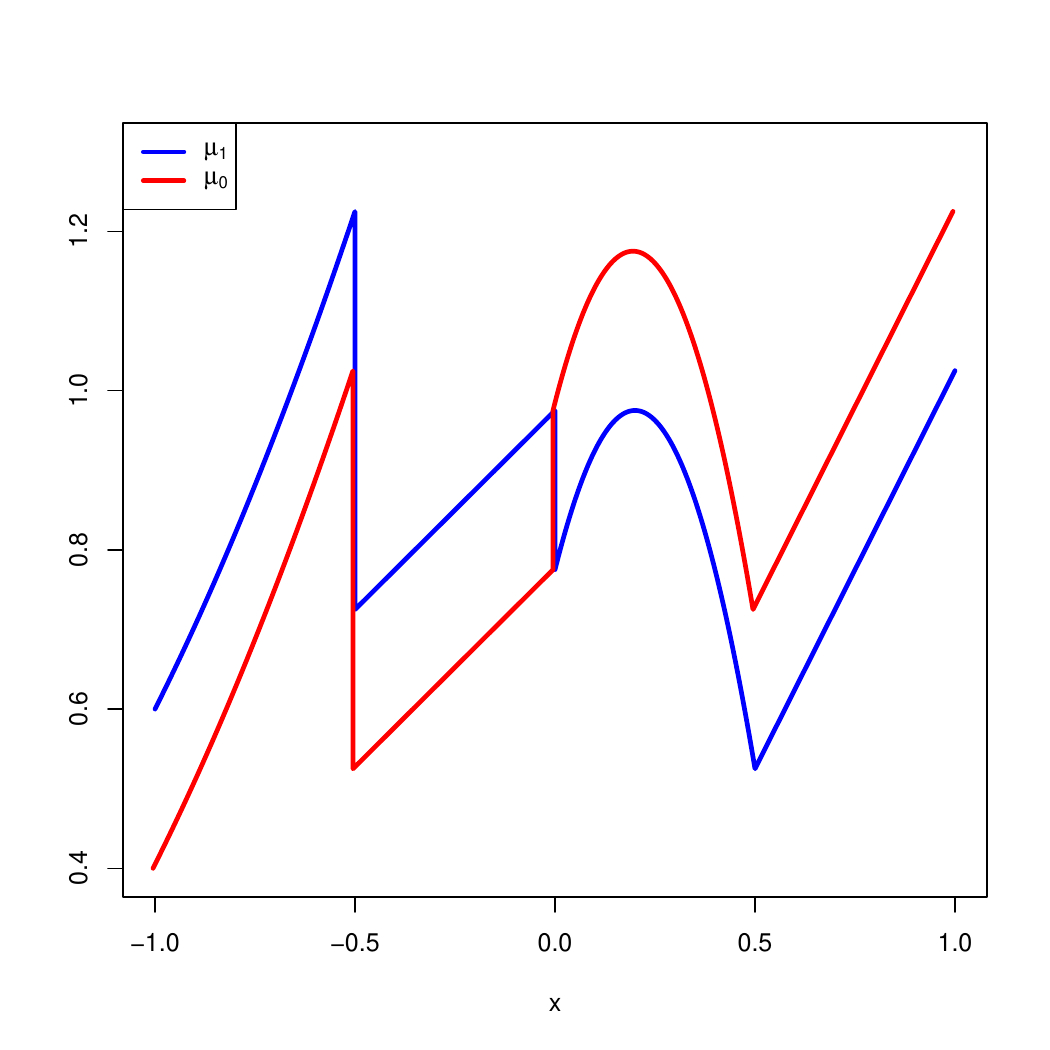}
\caption{Example regression functions $\mu_w(x)=\E[Y_i \mid X_i=x, W_i=w]$ for which (i) the average causal effect $\E[\mu_1(X_i)-\mu_0(X_i)]$ is zero (\eg if $X \sim \text{Unif}[-1,1]$), (ii) the individual regression functions $\mu_a$ are non-smooth and difficult to estimate accurately, and (iii) the CATE $\mu_1(x)-\mu_0(x)=-0.2 \times \text{sign}(x)$ is piecewise constant and  very simple.} \label{fig:cateplot}
\end{figure}

Much of causal inference has focused on simple summaries of causal effects, for example the average treatment effect, or average effect on the treated. Such global averages, however, can mask important  heterogeneity; 
for example, if half of the subjects benefit and half are harmed, the average effect can be zero (see, \eg  Figure \ref{fig:cateplot}).

On its own, effect heterogeneity is critical for scientific progress, as it can shed light on the mechanisms that determine why  some patients, but not others, respond to treatment. 
Effect heterogeneity is also central for targeting treatments or setting treatment policies, also known as \emph{policy learning}. In the toy example above, if we knew which half of patients benefited from treatment, we could treat only them, avoiding harm to the other half. There are even more possibilities when the treatment takes on more than two levels, such as drug dosage, or when the treatment can vary over time.

\subsubsection{Background}
For a binary treatment, effect heterogeneity is often quantified and explored via the 
conditional average treatment effect (CATE), defined as
$$ \E[Y_i(1) - Y_i(0) \mid X_i=x] , $$
for covariates $X_i$ (or possibly a subset of the available covariates).
This estimand is a
natural regression analogue of an average causal effect and plays a fundamental role in policy learning. 
Namely, in the binary treatment setting, let $\pi: \mathbb{X} \rightarrow \{0,1\}$ denote some policy for assigning treatment based on covariates $X_i$, and let $Y_i(\pi(X_i))$ denote the potential outcome for unit $i$ under policy $\pi(\cdot)$. The optimal treatment policy  for maximizing the mean outcome is
$$ \one\Big\{ \E[Y_i(1) - Y_i(0) \mid X_i=x] > 0 \Big\} , $$
\iiee, the policy that simply thresholds the CATE at zero and treats everyone whose CATE is positive. Note that the maximization in the above is over all possible policies $\pi: \mathbb{X} \rightarrow \{0,1\}$; however it is often of interest to only maximize over a smaller constrained class (possibly due to budget limits or interpretability), in which case the relationship to the CATE can be less clear. The field of policy learning is very large, spanning statistics, econometrics, machine learning, and more \citep{murphy2003optimal, hirano2009asymptotics, athey2021policy, dudik-offline-1}, with connections to other large literatures on bandit problems and reinforcement learning  \citep{bubeck-survey}. See Section \ref{section:experiments} for related connections to experiments. 

The CATE is fundamentally different from---and more difficult to estimate than---the ATE, analogous to how estimating a regression function is different from estimating a mean. 
For example, in an oracle setting where the potential outcome differences $Y_i(1)-Y_i(0)$ were actually observed, then the ATE could be estimated at parametric ($1/\sqrt{n}$) rates with just a sample average. However, estimating the CATE would require a regression of $Y_i(1)-Y_i(0)$ on covariates $X_i$; and classic no-free-lunch theorems from nonparametric regression \citep{stone1980optimal, gyorfi2006distribution} tell us that no non-trivial rates of convergence can be achieved without extra assumptions about this regression function (\eg smoothness or discreteness of regressors). 
Differences become even more stark outside the oracle setting, where potential outcomes must be estimated. 

In early forays towards estimating the CATE, parametric and semiparametric models were popular, where the CATE was assumed to follow some particular (partly) parametric form (\eg constant, or linear in a subset of covariates) \citep{robinson1988root}. More recently, methods using flexible nonparametric estimation and machine learning tools have become more popular, with a particular emphasis on double robustness \citep{athey2015recursive, foster2023orthogonal, kennedy2023towards, chernozhukov2018double, robins2008higher, athey2021policy, semenova2021debiased, van2005statistical}. 
Similarly, earlier work on policy learning typically focused on randomized experiments with known propensity scores or parametric models where conditional effects can be estimated at parametric $1/\sqrt{n}$ rates \citep{kitagawa2018should, hirano2009asymptotics}. Recent work has focused on the setting with unconfoundedness, allowing for flexible nonparametric models and for incorporating black-box machine learning tools \citep{athey2021policy, luedtke2020performance}. 

\subsubsection{Challenges}

Much of the literature on heterogeneous effects and policy learning has focused on expectations, mostly in smoothness models, with relatively simple data structures (\eg single time-point binary treatments). There are some exceptions of course,  including \citet{wang2018quantile, leqi2021median, bradic2019minimax, lewis2021double}, and others. However there is much to explore outside this initial case. 

\paragraph*{Beyond expectations.} Part of the focus on expectation-based contrasts in heterogeneous effect estimation stems from the connection to policy learning, and the fact that such contrasts are connected to mean-optimal treatment policies. However, these contrasts are often used in settings where explicit policy learning is not the main goal; thus there  are potentially many other measures of heterogeneity that could be useful. These include heterogeneity based on variances and other features of the entire distributions. Further, it can be fruitful to explore policies that are optimal for objective functions that depend on other aspects of the distributions beyond means, such as inequality measures.

\paragraph*{Beyond smoothness.} Although smoothness models (where some number of derivatives are assumed bounded) are a standard framework from nonparametric statistics, many different forms of structure can be considered, each with their own nuances and advantages/disadvantages. Models relying instead on sparsity, or bounded variation, or neural network structure are underexplored for heterogeneous effect estimation, for example. 

\paragraph*{New data structures.} Although studies with single time-point binary treatments are conceptually simple, and a natural starting point, real data are often much more complex, involving continuous or multivariate treatment options that may change over time, influenced by previous treatments and covariates that also change over time, for example. Further, covariates and outcomes can be high-dimensional and non-Euclidean (\eg\ graphs or images), there can be multiple studies from different sources, and subjects can be connected in complex networks \citep{ogburn2024causal, kurisu2024geodesic}. There is much to do exploring heterogeneous effect estimation and policy learning in these non-standard but increasingly common data structures. 
Finally, a long line of work starting with \citet{robins1986new} has focused on estimating optimal treatment regimes, that is, optimal dynamic policies; this is a much more challenging setting.

\paragraph*{Inference.} In nonparametric regression, there are fundamental challenges that come with inference (\eg\ constructing valid confidence bands) compared to estimation. These challenges are just as fundamental for the CATE, and so have led to a dearth of inferential versus estimation procedures. For example, optimizing mean squared error requires balancing squared bias and variance, but this complicates inference, since one is left with a non-trivial bias that does not shrink when scaled by the standard error; \citet{wasserman2006all} calls this ``the bias problem''. Further, although one can construct estimators that automatically adapt to unknown smoothness or other structure, there are strong negative results indicating how confidence bands cannot adapt in the same way \citep{low1997nonparametric, genovese2008adaptive}. These exact same issues arise for the CATE, and can be exacerbated due to the fact that the potential outcomes are not directly observed, and so there is additional interplay with nuisance estimation error. More work is needed to understand the role of these inferential challenges in causal-specific problems (\eg\ what are the implications for policy implementation), and to consider how some proposed fixes work for causal inference specifically (\eg\ bias reduction via debiasing or undersmoothing, settling for covering simpler CATE surrogates, or accepting weaker coverage guarantees).

\paragraph*{New optimization goals.}
As described above, the classic setup in policy learning is to estimate policies $\pi(\cdot)$ that maximize the mean outcome $\E[Y_i(\pi(X_i))]$, often under assumptions of no unmeasured confounding and overlap, for example. However, practitioners often care about more than mean outcomes, and also want to balance optimizing outcomes with, for example, implementing policies that assign treatments in fair ways, or respect constraints on treatment availability, or other constraints. Further, in practice the typical identifying assumptions may be violated, or the goal may be framed in terms of the joint distribution of potential outcomes, in which case one could  bound the optimal policy or identify a policy with minimal worst-case regret. Although there have been recent advances in this vein \citep{kallus2021minimax, ben2024policy}, there may be other important alternative approaches to explore, \eg\ in settings with distribution shift, where policies may be implemented in new populations that differ from that used in training.

\paragraph*{Implementation, translation, and beyond.}
As is true in many areas of causal inference, for heterogeneous effect estimation and policy learning there is also a gap between modern theory and methods, and practical implementation. More work is needed studying how to apply state-of-the-art methodology in practice to substantive research problems, and making statistical tools more available and accessible. 

Finally, one can construct heterogeneous treatment effect-based versions of many if not all of the challenges presented in this paper, to obtain other new and interesting challenges. For example, heterogeneous effects in complex experiments, with interference, combined with sensitivity analysis, from a discovery perspective, with automation, \emph{etc}., are all interesting and valuable areas to pursue. Compared to the versions of these problems involving more standard summary effects, versions focusing on heterogeneous effects and policy learning may come with substantial nuances and differences.

\subsection{Mediation and Causal Mechanisms}

\subsubsection{Motivation.}

Much work in causal inference has focused on quantifying treatments' effects on an outcome. Even gold-standard experiments ``reveal but do not explain causal relationships'' \citep{bullock2010yes}. However, it is often of interest to go beyond a black-box assessment of \emph{whether} treatment works, and to also learn \emph{how} it works; this is the goal underlying mediation analysis.

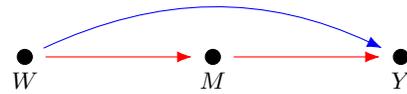
\begin{figure}
    \begin{tikzpicture}[
        >=stealth,
        node distance=2.0cm
        ]
        \node[observed, label=below:{\(W\)}] (2) at (2.5,0) {};
        \node[observed, label=below:{\(M\)}] (3) at (5,0) {};
        \node[observed, label=below:{\(Y\)}] (4) at (7.5,0) {};
        
        \draw[->, notouch, red] (2) -- (3);
        \draw[->, notouch, blue, bend left=25] (2) to (4);
        \draw[->, notouch, red] (3) -- (4);
        
    \end{tikzpicture}
\caption{Directed acyclic graph for the standard mediation setup. $M$ is a mediator variable on the path between treatment $W$ and outcome $Y$. The blue path represents the \emph{direct effect} of $W$, while the red path shows the \emph{indirect effect} of $W$ \emph{through} $M$.}
\label{fig:meddag}
\end{figure}

Figure \ref{fig:meddag} shows a directed acyclic graph in a standard mediation setting. Here the goal may be not only to understand how treatment $W$ affects outcome $Y$, but also how much $W$ affects $Y$ \emph{directly} versus \emph{indirectly} through a mediator $M$. 
Note the differences between Figure \ref{fig:iv} with a DAG for an instrumental variables setting. Although in both cases we have a three variable setting with a causal path from the left to the right, the estimands and assumptions are quite different. In the instrumental variables case the direct link between the variable on the left (the instrument $Z$ in Figure \ref{fig:iv}) and the outcome $Y$ is absent, and there is an unobserved confounder affecting both the variable in the middle and the outcome. Understanding such mediation effects can be useful for many purposes. 
To make this more specific,
consider \citet{chen2019impact}. They study the effect of having a younger brother on a first child's education, surmising there may be a direct effect by taking away resources that are spent on the brother instead, and an indirect effect that having a brother may lead to fewer siblings overall (if there is a general desire for male children). Understanding these effects may help design educational interventions that affect children's outcomes.

Alternatively, furthering our understanding of the scientific mechanism of a treatment could be important as an end in and of itself. 

\subsubsection{Background.}

Mediation analysis has its origins in linear structural equation models \citep{wright1921correlation} and was popularized by \citet{baron1986moderator} in a linear regression setting.
\citet{robins1992identifiability} and \citet{pearl2001direct} later formalized mediation analysis within an explicit, nonparametric causal inference framework. Since then, mediation analysis and adjacent areas have become a rich and active area of research, with important applications in the health and social sciences, and emerging areas such as algorithmic fairness and policy evaluation.
See \cite{vanderweele2015explanation} for more of an overview.

\paragraph*{Notation and common estimands.}

One of the earliest and most widely used estimands in mediation analysis is the \emph{controlled direct effect} (CDE). 
Let $Y_i(w, m)$ denote the potential outcome that would be observed if both the treatment were set to $W_i = w$ and the mediator to $M_i = m$. The CDE of changing the treatment from $w'$ to $w$, while holding the mediator fixed at $m$, is given by:
$$
\text{CDE}_m(w, w') := \mathbb{E}\left[ Y_i(w, m) - Y_i(w', m) \right].
$$
This estimand captures the idealized setting of controlled experimentation, in which the scientist isolates the direct effect of the treatment by manipulating it while holding potentially mediating variables constant.

While conceptually straightforward, controlled effects can present limitations in answering mediation questions.
Importantly, CDEs fail to capture a core objective of mediation analysis, which is indirect effects, \iiee the effect
of the treatment on the outcome that occurs only because the treatment influences the mediator. One can consider controlled effects of the mediator changing from $m'$ to $m$, while holding the treatment fixed at $w$, \iiee 
$$
\mathbb{E}\left[ Y_i(w, m) - Y_i(w, m') \right],
$$
but this can be non-zero even when the treatment does not affect the mediator, and so is not a proper indirect effect.

These considerations lead to the ideas of \emph{natural} direct and indirect effects. Defining these objects require not only considering counterfactual mediators under different treatments, $M_i(w)$, but also nested counterfactuals, such as $Y_i(w, M_i(w'))$---that is, the potential outcome had the treatment been set to $w$, while letting the mediator take its natural value had the treatment been set to $w'$.  Then, the natural direct effect of changing the treatment from $w'$ to $w$, while letting the mediator take its natural level under $w$ is  defined as
\[
\text{NDE}_{w}(w, w') := \E[ Y_i(w,M_i(w)) - Y_i(w',M_i(w))].
\]
Similarly, the natural indirect effect of holding the treatment fixed at $w'$, while allowing the mediator to vary as if we had changed the treatment from $w'$ to $w$, is
$$
\text{NIE}_{w'}(w,w') := \E[ Y_i(w', M_i(w)) - Y_i(w', M_i(w')) ].
$$
Importantly, now the average treatment effect can be decomposed into a sum of the natural direct and indirect effects,
\begin{align*} 
\E[Y_i(w)&-Y_i(w')] = \E[Y_i(w,M_i(w))-Y_i(w',M_i(w'))] \\
&= \E[Y_i(w,M_i(w)) - Y_i(w',M_i(w))] \\
& \hspace{.4in} + \E[ Y_i(w',M_i(w)) - Y_i(w',M_i(w'))] \\
&= NDE_{w}(w, w') + NIE_{w'}(w, w').
\end{align*}
Other decompositions of the ATE are possible, accounting for interactions between the treatment and the mediator  [\eg \citealp{vanderweele2014unification}].

Controlled and natural effects can both be useful in characterizing mediation, depending on the setting. Controlled effects can be viewed as representing direct manipulation, while natural effects can be more representative of underlying causal mechanisms \citep{pearl2001direct}.

There are also randomized interventional versions of natural effects \citep{robins2003semantics, didelez2006direct, van2008direct}, where the mediator is not set to its actual values $M_i(w)$ or $M_i(w')$, but instead to random draws from the corresponding conditional distributions, \eg to some $G_i(w) \sim dP(M_i(w) \mid X_i)$.

\paragraph*{Identification \& estimation.}

Since controlled effects are solely effects of a joint exposure $(W,M)$, their identification follows as in other causal settings, \eg via no unmeasured confounding in the form $(W,M) \perp\!\!\!\!\perp  Y(w,m) \mid X$, along with consistency and positivity. 
For natural effects, identification is somewhat more involved. The terms in natural effects involving counterfactuals of the form $Y_i(w,M_i(w))=Y_i(w)$ can be identified as usual, since this is just an effect of setting $W$. However, both natural direct and indirect effects also involve counterfactuals of the form $Y_i(w,M_i(w'))$, where the treatment is set to some $w$ but the mediator is set to its counterfactual under a different treatment $w'$. The quantity $\E[Y_i(w,M_i(w')) \mid X]$ can be identified as
$$ \int \E[Y \mid X, W=w, M=m] \ dP(m \mid X, W=w')  $$
under the same consistency, positivity, and no unmeasured confounding assumptions used for direct effects, together with two additional assumptions. The first is no unmeasured confounding for the effect of treatment on mediator, $W \perp\!\!\!\!\perp M(w') \mid X$, which is analogous to standard assumptions. 
The second is known as \emph{cross-world exchangeability}, $M(w') \perp\!\!\!\!\perp Y(w,m) \mid X$, which is a qualitatively different assumption 
since it concerns counterfactuals under different treatments $w$ and $w'$. 
In particular, cross-world exchangeability cannot be enforced experimentally by randomization and cannot be falsified \citep{miles2023causal}, although the randomized interventional versions of these effects can be identified without cross-world conditions.

After identification, estimation can proceed much as in other causal problems, for example via outcome modeling, inverse propensity score weighting, matching, semiparametric doubly robust methods, \emph{etc}. Semiparametric theory was developed by \citet{tchetgen2012semiparametric}, who showed corresponding estimators are ``triply robust''. Although this sounds more favorable than ``doubly robust'', it is actually less so---in mediation there are three nuisances instead of two (usual treatment and outcomes distributions, as well as the mediator distribution), and two of the three nuisances need to be estimated consistently (rather than one of two in the doubly robust case).

\subsubsection{Challenges}

\paragraph*{Designing experiments for mediation analysis.}
While most mediation analysis are based on observational studies, experimental designs can provide stronger causal evidence of the underlying mechanisms. 
An interesting methodological challenge is the design of experiments that can uncover mediation effects.
One promising approach is sequential randomization. 
For instance, participants might first be randomized to a smoking cessation program, after which their motivation is measured, and subsequently they can be randomized again to receive a motivational enhancement. 
This sequential assignment can help distinguish effects on the mediator from those on the outcome.
Beyond sequential randomization, factorial designs allow simultaneous manipulation of both the treatment and the mediator, provided direct manipulation of the mediator is feasible. 
Encouragement designs can be employed when the mediator cannot be directly manipulated, by randomizing an encouragement or incentive intended to shift the mediator. 
These experimental strategies can be developed depending on the context, enabling stronger causal claims about mediation pathways.

\paragraph*{Representing complex mediators.}
Another exciting challenge arises when dealing with high-dimensional, unstructured mediators, such as those present in multi-omics, digital trace, or text data. In these contexts, there is often no clear mechanistic theory that specifies the mediators in advance; instead, the mediators must be learned, summarized, and interpreted directly from the data at hand. This raises interesting questions, including how to define appropriate causal estimands and articulate meaningful identification assumptions. Addressing this gap also presents opportunities to leverage advances in representation learning and related areas \citep{scholkopf2021toward} to detect and characterize mediators in a data-driven manner. The overarching goal is to develop causal models that can succinctly capture the effects of such mediators, enabling inference and interpretability in settings where there are many candidate mediating pathways both diffuse and overlapping in nature.

\paragraph*{Decomposing causal pathways.} Another relevant direction is the decomposition of causal effects, moving beyond traditional direct and indirect effects to assess, for example, disparities in health care access driven by gender or race \citep[\eg][]{jackson2021meaningful}.  This line of research is closely connected to issues of fairness and opportunity. However, these questions are particularly challenging, as causal effects and social constructs like race are deeply intertwined and often difficult to disentangle, even at a conceptual level. Nevertheless, this area is crucial for both policy and fairness considerations, making it essential to develop clear frameworks and language to rigorously discuss these complexities \citep{howe2022recommendations}.  

\paragraph*{Studying time-varying mediation.}
From a dynamic viewpoint, it is important to understand how direct and indirect effects evolve over time \citep{vanderweele2017mediation}. A compelling emerging application is in digital health interventions, where app usage influences behavior change, which, in turn, affects health outcomes.
Among others, these settings are challenging because of complex direct and indirect effects through multiple time points. Current methodological approaches include G-methods and Marginal Structural Models (MSMs) \citep[\eg][]{aalen2020time, mittinty2020longitudinal, diaz2023efficient}. 
Theoretical challenges involve defining estimands that are both meaningful and based on realistic assumptions, as well as decomposing effects through multiple mediator pathways. 
Practical challenges include sparse data across numerous causal pathways, issues of dropout and missing data, and the need to appropriately model time trends.

\paragraph*{Generalizing mediation effect estimates.}
Often, just as it is crucial to understand the effects of an intervention in a target population, so too is uncovering the mechanisms underlying those effects in the population of interest.
This requires generalizing or transporting estimates of direct and indirect (mediated) effects from a study population to a target population that may differ in covariate distributions, effect modifiers, and potentially in mediators themselves.
Key challenges in this context include unmeasured confounding, the need for robust estimation approaches, and reliance on strong assumptions involving cross-world counterfactuals.

\paragraph*{Conducting mediation analysis in practice.}
In a guide for applied researchers, \citet{schuler2024practical} provide a thoughtful discussion of practical challenges encountered in mediation analysis. 
Among these, they highlight the importance of selecting mediation effects that accurately reflect the scientific question of interest, assessing the validity of the underlying assumptions of no unmeasured confounding, and addressing measurement error in the mediator.
They also underscore the importance of comprehensive and transparent reporting of the results of mediation analyses.
Addressing these challenges can greatly foster mediation analysis in practice.

\subsection{Optimality and Minimaxity}

\subsubsection{Motivation}
Distinct issues arise in causal effect estimation, subsequent to identification. In particular, researchers often want to use flexible methods that do not rely on strong modeling assumptions. This raises important questions, such as: What kind of statistical structure are we willing to assume? How do we build estimators that work well under these assumptions? What is the \emph{best possible  performance} we could hope to achieve under our assumed structure? The minimax framework is a natural and powerful one for considering these kinds of questions, with a long history in statistics going back half a century \citep{lecam1973convergence,le1986asymptotic, stone1980optimal, tsybakov2008introduction}. 

Although minimax optimality is well understood in some settings, such as regression and density estimation, relatively little is known when it comes to causal inference problems. There is a pressing need to address this gap: it means that in causal inference we often {do not know} whether we could be making better, more efficient use of the data, whether there are better estimators out there waiting to be discovered, or if what we have is the best we could ever get. This can lead to a proliferation of different new methods, with little  clarity about how to compare or benchmark them. And this lack of clarity is uniquely concerning in causal settings, where empirical benchmarking can be inherently limited due to lack of a ground truth \citep{holland1986statistics}. In contrast, the minimax framework provides a path forward for \emph{theoretical benchmarking} in causal inference---even if this path is rife with technical challenges.

\subsubsection{Background}

We first outline the basic setup for minimax estimation; 
a very useful review can be found in \citet{tsybakov2008introduction}. We start out with some desired statistical target, denoted $\psi$. This could be a function, like a conditional expectation $\E[Y \mid X=x]$, or a real-valued parameter (\iiee functional), like the causally motivated quantity $\int \E[Y_i \mid X_i=x, W_i=1 \ dP(x)$, which equals $\E[Y_i(1)]$ under no unmeasured confounding assumptions. Next, we decide what we want to assume statistically, which amounts to settling on a model $P\in\mathcal{P}$, {\it i.e.}, a set of distributions we assume contains the truth. Common examples include  parametric models, where distributions only differ up to finitely many parameters, or nonparametric models, where distributional components are infinite-dimensional and only assumed to have some smoothness, sparsity, bounded variation, \emph{etc}. 

Given an assumed model  $\mathcal{P}$, the minimax rate for estimating the statistical target $\psi=\psi(P)$ with loss $\ell$ is given by
$$ R_n = \inf_{\widehat\psi} \sup_{P \in \mathcal{P}} \E_P\left\{ \ell\left( \widehat\psi, \psi(P) \right) \right\} $$ 
where the infimum is over all possible estimators, \iiee all possible functions of the data. In words, this is the best possible (worst-case) estimation error, across all possible estimators. In practice, minimax rates are typically characterized in two steps: first we find a lower bound, showing that $ R_n \geq C r_n$ for some constant $C$ and sequence $r_n$, and second we find an upper bound, showing that $R_n \leq C' r_n$ for some possibly different constant $C'$. Then we can conclude the minimax rate is $r_n$. A lower bound on the minimax rate is a powerful result: it says that no estimator can have risk (\iiee expected loss) uniformly smaller than $C r_n$, no matter how creatively it is constructed, or how much computation it involves. 

 Minimax rates thus have crucial implications, both practical and theoretical. First, they give a precise benchmark for the best possible performance of a statistical task. If some estimator's risk does not match a minimax lower bound, then more work is needed---either at improving the estimator (or its analysis), or improving the bound. Alternatively, if some estimator's risk \emph{does} match a lower bound, then it cannot be improved (at least in terms of rates), without adding or changing assumptions. Second, minimax rates precisely characterize the fundamental limits of estimation in a given problem. This allows us to compare different tasks, and order them in terms of their inherent statistical difficulty in interesting ways. 

\begin{figure}[!ht] 
\centering
\includegraphics[width=3.4in]{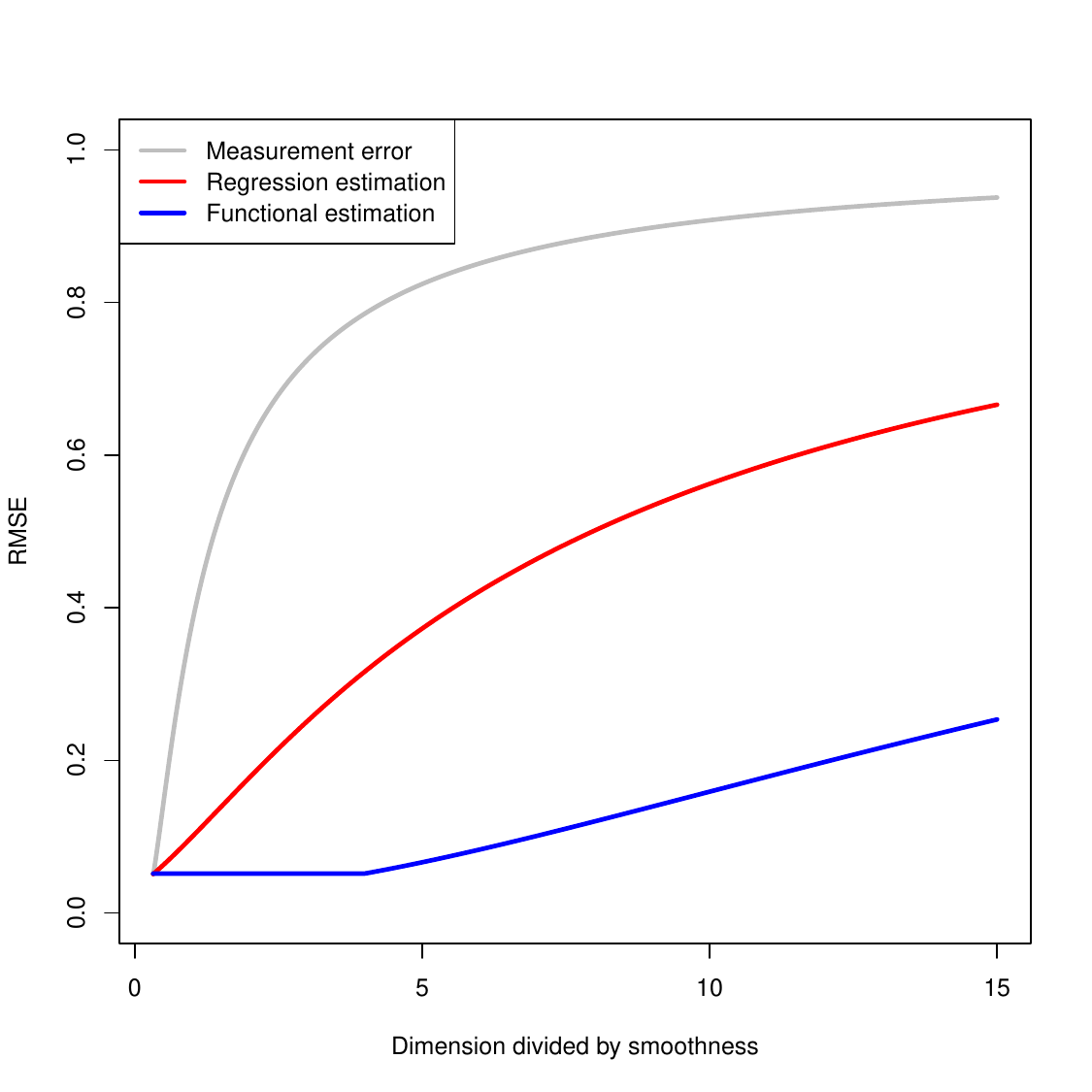}
\caption{Minimax rates for three classical estimation tasks, as a function of covariate dimension $d$ scaled by smoothness $s$: functional estimation (\eg\ estimating the ATE in causal inference); regression estimation (\eg\ the CATE); and density estimation with measurement error (\eg the counterfactual density when outcomes are measured with error). In causal analogues of these problems the rates further depend on the complexity of nuisance functions (\eg\ propensity scores, outcome regressions).} \label{fig:minimaxplot}
\end{figure}

In some settings, much is known about minimax rates. For example, standard parametric models have been studied for decades and precise local rates are available \citep{van2000asymptotic}. For smooth nonparametric models, where functions are only assumed to have $s$ bounded derivatives,  minimax rates for $d$-dimensional nonparametric regression and density estimation are of the form
$$ n^{-1/(2+d/s)} $$
in terms of RMSE \citep{stone1980optimal}; see Figure \ref{fig:minimaxplot} for an illustration. This is like the parametric rate $1/\sqrt{n}$, but with an added penalty for how large the dimension $d$ is relative to the smoothness $s$. For quadratic functionals in smooth nonparametric models, such as the expected density $\E\{p(X_i)\} = \int p(x)^2 \ dx$, the minimax rate has been shown to be
$$ \max\left( \frac{1}{\sqrt{n}} , \  n^{-1/(1+d/4s)} \right) $$
\citep{birge1995estimation, laurent1996efficient}; see Figure \ref{fig:minimaxplot} for an illustration (and note the ``elbow'' shape). This shows how estimating functionals of a density can be statistically easier than estimating the whole density itself: the rate is parametric when $s \geq d/4$, and otherwise the penalty driving the rate away from parametric is weaker than for density estimation, depending on $d/4s$ instead of $d/s$. For smooth density estimation with Gaussian measurement error, the minimax rate is
$$ (\log n)^{-s/2} $$
\citep{carroll1988optimal}, which is very slow, indicating the inherent difficulty of the problem; see Figure \ref{fig:minimaxplot} for an illustration. Thus in this problem one either needs to live with large estimation errors or else add more structure/assumptions. 

Although minimax optimality has a relatively long history in the aforementioned settings,  far less is known when it comes to causal inference. The causal inference setting where most is known about minimax rates is in estimation of sufficiently smooth (\eg pathwise differentiable) target parameters, when they can be estimated at parametric rates. In this setting, there are precise convolution theorems for regular estimators, and local asymptotic minimax rates available \citep{hajek1972local, bickel1993efficient}. However, these rates are not always attainable, \eg if covariate dimension is too large or nuisance function smoothness is too low.
Thus, over the last decade or so, Robins and colleagues have developed extensive results for pathwise differentiable functionals \emph{even when parametric rates are not achievable},  in smooth nonparametric models \citep{robins2008higher,robins2017minimax}.

\subsubsection{Challenges}

\paragraph*{New parameters.} 
Even for the most basic causal inference parameter  $\psi(P)=\int \E[Y_i \mid X_i=x, W_i=1] \ dP(x)$ (and even in classical smoothness models), there are still open and unsolved problems regarding the role of the covariate distribution. When one moves to other more complicated parameters, many more open problems arise. For example, stochastic intervention effects are by now of wide interest \citep{kennedy2019nonparametric, munoz2012population, haneuse2013estimation}, but optimality has not been explored outside the $\sqrt{n}$ setting. Similarly,  optimality for time-varying treatment effects outside the $\sqrt{n}$ regime is entirely open, regardless of the model setup. \citet{kennedy2024minimax} recently characterized minimax rates for heterogeneous effect estimation in smoothness models, but there is little work outside this regime. For similarly complex parameters like dose-response curves \citep{kennedy2017non} or counterfactual densities \citep{kim2018causal},  little is known.

\paragraph*{Adaptivity.} Most available results and methods in the context of optimality have been developed in settings where relevant complexity parameters, such as the amount of smoothness, are known (with few exceptions, {\it  e.g., }\citet{mukherjee2015lepski}). Adaptivity is a fundamental aspect of nonparametric estimation \citep{donoho1995adapting, efromovich1996sharp, low1997nonparametric}, but it has not been explored in detail in causal inference problems. Thus there are many fundamental open questions of how and to what extent it is possible to adapt to unknown smoothness, or more generally, across sets of different function classes. 

\paragraph*{New models.} 
The  large majority of work on minimax optimality in causal inference considers smoothness models, where nuisance functions are assumed to have some number of bounded derivatives \citep{robins2009semiparametric, kennedy2024minimax}. (Recall functions with more bounded derivatives are less complex, and easier to estimate.) Smoothness models are a natural starting point: they are the prototypical nonparametric function class, for which classic kernel/series methods can work well, and which are infinite-dimensional but also structured enough to allow for fast convergence rates, depending on dimension/smoothness. But many other models are also of  interest (and potentially more practical), including high-dimensional models \citep{bradic2019minimax, liu2023root, zeng2024causal}, bounded variation models \citep{van2017generally}, structure-agnostic models \citep{balakrishnan2023fundamental, jin2024structure}, neural network models \citep{farrell2021deep}, and more. 

\paragraph*{Implementation.} In the relatively few settings where estimators are available that can improve on doubly robust-style methods (for example, using higher-order methods \citep{robins2008higher, robins2017minimax}, undersmoothing \citep{newey2018cross, mcgrath2022nuisance, mcclean2024double}, or exploiting linearity in high-dimensions \citep{athey2018approximate,smucler2019unifying,wang2024debiased, liu2023root}), the estimators are often computationally intensive and require careful tuning. There are many opportunities to make these methods more practical, automatic, and user-friendly.

\paragraph*{New frameworks.} There are variations on the minimax optimality framework described above, explorations of which present new opportunities that may be particularly important in causal inference problems. For example, in many statistical problems it is common to use straightforward losses, \eg squared error $\ell(\widehat\psi,\psi_P)=(\widehat\psi-\psi_P)^2$, or integrated $L_2$ or supremum errors  for functions. However, in causal problems it may be worthwhile to consider new losses, \eg that connect to the threat of unmeasured confounding, or are derived directly from feedback from policymakers in some way. 

Similarly, expected loss may not always be most meaningful; recent work has explored minimax quantiles or confidence interval lengths, for example \citep{ma2024high}. Another variant is to consider local minimax optimality, where the supremum in $R_n$ is not over the whole model $\mathcal{P}$, but instead over a subset near some specific distribution $P$, \iiee over all distributions $Q$ such that $d(Q,P)<\epsilon$ for some distance $d$ \citep{van2000asymptotic, balakrishnan2019hypothesis}. Local minimax rates are more nuanced and can illustrate how fundamental limits can change with a particular distributions $P$. Beyond these variations, it may be worthwhile developing other new frameworks for optimality, which depart even further from the minimax setup.

\subsection{Sensitivity Analysis and Robustness}
\label{sec:sensitivity}

\subsubsection{Motivation.} 
Canonical methods used for drawing causal inferences from observational data almost by definition rely on strong, often untestable assumptions about the data generating process. For example, estimating causal effects through covariate adjustment assumes no unobserved confounders between the treatment and the outcome; instrumental variable methods require both an assumption of no unmeasured confounders between the instrument and the outcome, and an exclusion restriction that rules out direct effects of the instrument on the outcome; difference-in-differences designs assume that, in the absence of treatment, the average outcome for the treatment and control groups would have evolved in parallel over time.

While such assumptions may provide a useful benchmark and starting point for causal analysis, they are often unlikely to hold exactly in real-world settings. This raises several questions: how sensitive are causal inferences to violations of such key assumptions?  Or, how large would deviations from such assumptions need to be to substantively change the main results of a study?  And how plausible are such deviations in a given context? 
Probing the \emph{robustness} of causal claims to departures from causal assumptions is an important and active area of research known as \emph{sensitivity analysis}.

\subsubsection{Background.} 
The first—and arguably most successful—empirical application of sensitivity analysis is the seminal study by \cite{cornfield1959smoking}, which assessed the causal nature of the association between cigarette smoking and lung cancer. In the 1950s, several observational studies found that smokers were substantially more likely than non-smokers to develop lung cancer  \citep[\eg][]{doll1950smoking}. At the time, however, Sir Ronald Fisher and other prominent statisticians argued that, in the absence of experimental evidence, unobserved confoudners, such as an individual's genotype, could provide an alternative explanation for the observed association \citep{fisher1957dangers,fisher1958cigarettes}.

The now-classic sensitivity analysis by \citet{cornfield1959smoking} demonstrated that, even if such an unobserved confounder existed, this hypothetical ``smoking gene'' would need to be at least nine times more prevalent in smokers than in non-smokers to fully account for the observed association---something deemed scientifically implausible by most experts then and now. Absent such a strong relation, they concluded that a causal link between cigarette smoking and lung cancer is still necessary to explain the observed association, even allowing for plausible amounts of unmeasured confounding. Building on Cornfield's seminal work, an extensive body of literature has developed quantitative frameworks for sensitivity analysis. Here we present a non-exhaustive sample of methods developed to perform sensitivity analysis to unobserved confounders.

A common thread of many approaches is to assume that unconfoundedness holds, but  only after adjusting both for observed covariates $X_i$ as well as for an unobserved covariate $U_i$, that is,
$$
 W_i \perp\!\!\!\!\perp \Bigl(Y_i(1), Y_i(0)\Bigr) \  \Bigl|\ X_i,U_i.
$$
Given this augmented unconfoundedness assumption, researchers then impose limits on the association between the unobserved confounder and the treatment, or on the association between the unobserved confounder and the outcome, or on both. Alternatively, researchers can also compute the minimal strength of association of $U_i$ such that it would invalidate the main conclusions of the original study.

One of the earliest works in this area is given by \citeauthor{rosenbaum1983assessing} [\citeyear{rosenbaum1983assessing}], who posited a parametric model for the relationship between potential outcomes and both observed and unobserved confounders, along with a parametric model for the propensity score as a function of these (observed and unobserved) confounders. Given specific values for the parameters attached to $U_i$ governing these relationships, they then examined how variations in these parameters affect the estimated average treatment effect.  \citet{imbens2003} and \citet{cinelli2020making} extended this approach by linking the range of values for the sensitivity parameters to estimable associations given observed confounders. Other related approaches can be found in \cite{arah2011}, \cite{dorie2016flexible}, \cite{altonji2005selection}, and \cite{veitch:neurips2020}.

For binary treatments, a common approach is to limit solely the strength of confounding by placing bounds on the odds ratio of the treatment assignment distribution, conditioning and not conditioning on potential outcomes. This model has been extensively developed by \citeauthor{rosenbaum1987sensitivity} [\citeyear{rosenbaum1987sensitivity, rosenbaum2009amplification}] and more recently by \cite{tan:jasa2006,  yadlowsky:arxiv2018, kallus2018confounding, zhao2019sensitivity, jesson2021quantifying, dorn2023sharp, dorn2024doubly}. Alternative approaches for binary treatments include \cite{masten:e2018}, who place bounds on the difference between these distributions, and \citet{bonvini2022sensitivity}, who limits the fraction of units for which treatment assignment deviates from being as good as random after conditioning on observed confounders. Closely related to the original result of \cite{cornfield1959smoking}, \cite{ding2016sensitivity} derive general bounds for the causal risk-ratio, with sensitivity parameters also expressed in terms of risk-ratios relating the confounders with the treatment and the outcome.

Other approaches for sensitivity analysis  directly specify a ``tilting'' or ``bias'' function that relates the conditional distribution of the outcome under treatment (or control) between treated and control units, or posits the magnitude of the difference of mean (or other contrasts of) potential outcomes. Earlier work in this area dates back to \cite{robins1999association,robins2000sensitivity}, \cite{brumback2004sensitivity}, and \cite{diaz2013sensitivity}, with recent work by \cite{franks2020flexible} and \cite{scharfstein2021semiparametric}.  In another vein, \citet{cinelli2020making,cinelli2024omitted} derive omitted variable bias formulas for linear regression coefficients that depend on partial $R^2$ values describing how much residual variation unobserved confounders explain of the treatment and of the outcome.  This omitted variable bias approach has been generalized to fully nonparametric settings in \citet{chernozhukov2022long}, along with doubly robust estimators, for a broad class of common causal parameters, including average treatment effects for binary treatments and average causal derivatives for continuous treatments.

\subsubsection{Challenges.} 

The sensitivity analysis literature presents somewhat of a conundrum. On the one hand, there is the canonical \citet{cornfield1959smoking} study, which  has been extremely successful.
There is also a clear sense that decision makers using causal analyses value information about the robustness of these estimates that goes
beyond point estimates and standard errors. On the other hand---and despite the impressive amount of  methodological work and the general consensus that causal analyses should be accompanied by some  analyses to assess the sensitivity to the fundamental assumptions that support them---such methods have not yet been widely adopted in the empirical literature.  We now discuss some practical and theoretical open challenges of sensitivity analysis.

\paragraph*{Robustness metrics.}

When traditional assumptions are met, there seems to be a rough agreement as to which statistics should be reported in a causal inference study: (i)~a point estimate, reflecting our ``best'' guess of the target of inference; and (ii)~a standard error, or confidence/credible interval, reflecting a measure of \emph{statistical uncertainty} surrounding the point estimate. However, these statistics do not necessarily reveal how sensitive conclusions are to deviations from the very assumptions that justified them in the first place.  
An important line of research is to devise statistics that can be routinely reported to quickly summarize how sensitive (or robust) a result is to \emph{systematic} biases. Examples include the E-value of \cite{ding2016sensitivity} and  the robustness value of \cite{cinelli2020making}. A closely related idea is the design of observational studies with such robustness in mind \citep{rosenbaum2004design}.

\begin{figure}[t]
    \centering
    \includegraphics[scale=0.75]{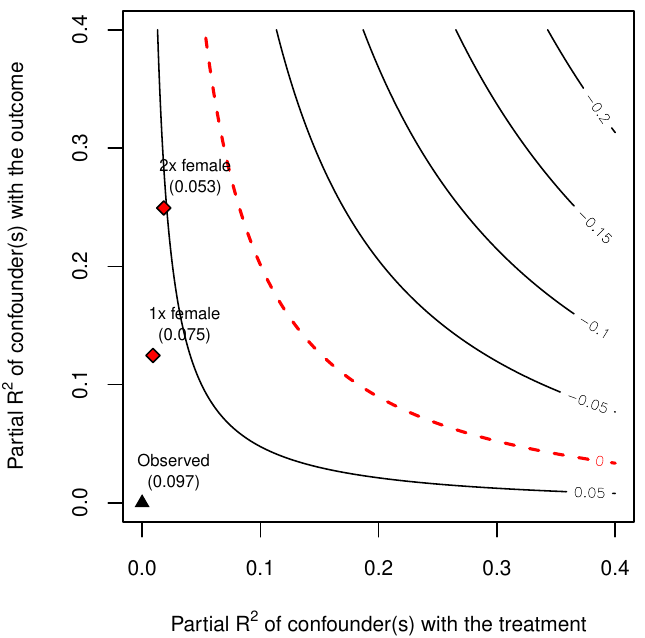}
    \caption{Example of a sensitivity contour plot with benchmarks \citep{cinelli2024sensemakr}. The horizontal axis describes the strength of association of unobserved confounders with the treatment, while the vertical axis describes their association with the outcome. The black triangle shows the original treatment effect estimate, assuming no confounding.  Contour lines show a bias-adjusted estimate, assuming confounding with the posited strength, and a direction of the bias that brings the estimate towards zero.  The red diamonds show bounds on the strength of unobserved confounders if they were as strong as the observed covariate ``female.'' In this example, the treatment effect estimate is still robust to unobserved confounding once or twice as strong as ``female.''}
    \label{fig:sensitivity-contour}
\end{figure}

\paragraph*{Connections to domain knowledge.}  One of the most challenging aspects of a sensitivity analysis is assessing the plausibility of violations that have been shown to be problematic. This usually requires making direct plausibility judgments on the magnitude of biases. Researchers, however, may leverage other types of domain knowledge to bound the strength of confounding. For example, \cite{imbens2003}, \cite{oster2019unobservable}, and \cite{cinelli2020making}, among others, propose comparing the relative strength of unobserved variables with the strength of observed variables. An example is provided in Figure~\ref{fig:sensitivity-contour}. It remains an open challenge how to meaningfully exploit such types of domain knowledge for sensitivity analysis.

\paragraph*{Other types of biases.} While much of the sensitivity analysis literature has focused on biases due to unmeasured confounding, other types of biases may also pose substantial threats to valid causal inferences. For examples, the problems of sample selection, attrition, missing data, measurement error, or interference plague not only observational studies, but also randomized experiments. Performing sensitivity analysis to these types of biases, and perhaps handling all of them simultaneously, is an important  open challenge.

\paragraph*{Sharpness results.} We say that a bound generated by a sensitivity analysis is sharp when it is the tightest possible bound given the sensitivity model and observed data. Obtaining sharpness results can often be difficult. For example, only recently have we obtained sharpness results for the sensitivity models of \citet{tan:jasa2006} and \cite{ding2016sensitivity}; see \citet{dorn2023sharp} and \citet{sjolander2024sharp}.

\paragraph*{Change in culture.} Current publishing practices may discourage researchers from questioning their own results. Thus, beyond technical challenges, advancing the adoption of sensitivity analysis requires a cultural shift on how researchers engage with imperfect assumptions and data. Rather than a threat to publication, sensitivity analysis should be seen as a tool for strengthening research, by drawing more robust and credible inferences under imperfect settings.

\subsection{Reliable and Scalable Causal Discovery}
\label{subsec:causaldisco}

\subsubsection{Motivation.} In most of the current causal inference literature we assume that we know at least qualitatively the causal relations between the treatment, the outcome and the covariates. However, in many real-world settings we do not know how the variables of interest relate to each other. For example, we might not even know which of the variables is a ``treatment'' and which an ``outcome.'' Additionally, it might be impractical or unethical to intervene on any individual variable. This is common in many disciplines, such as the biological, climate, and nutrition sciences. This challenge pertains to \emph{causal discovery} \citep{spirtes2001causation, glymour2019review}, \iiee learning causal relations between the variables of interest from data.

For example, in system biology scientists are interested in discovering how cells communicate with each other and how the signals they exchange get decoded in each cell through \emph{protein signalling networks}, networks of proteins that interact with each other \citep{sachs2005causal}. In this setting, we generally do not know which protein affects another, and our goal is to learn the causal relations between these variables from data, in order to be able to predict and control how certain signals affect a cell. This is particularly important in healthcare applications, \eg in cancer research, where cancerous cells do not respond to signals that would usually stop their growth.  

Learning causal relations from data, especially observational data, is an extremely challenging task: correlation and causation are not the same.
So, how can we actually learn such relationships? Intuitively, if we can assume that any causal relations between variables will make the affected variables associated each other, we can leverage patterns of association in the variables of interests to solve the inverse problem and narrow down the possible causal structures between them.

\subsubsection{Background.}
Reichenbach's principle of common causes \citep{Reichenbach1956-REITDO-2} provides a first example of an assumption that relates dependence and causation. According to this principle, if two variables $X$ and $Y$ are dependent, then either $X$ causes $Y$, $Y$ causes $X$ or there exists another variable $Z$ that causes both $X$ and $Y$. In this principle, saying $X$ ``causes'' $Y$ does not necessarily mean a direct causal effect, but possibly also an indirect causal effect. We can often additionally assume \emph{causal faithfulness} \citep{spirtes2001causation}, which in this case means that if two variables $X$ and $Y$ are independent, then none of these cases can happen.

\begin{figure}
    \centering
    \begin{tikzpicture}[>=stealth,node distance=1cm]

        \node[observed,label=above:{\(X\)}] (x1) at (0,0) {};
        \node[observed,label=above:{\(Y\)}] (y1) at (1.2,0) {};
        \node[observed,label=below:{\(Z\)}] (z1) at ($(x1)!0.5!(y1)+(0,-0.9)$) {};
        \draw[->,notouch] (x1) -- (z1);
        \draw[->,notouch] (z1) -- (y1);
        \node[align=center,scale=0.75] at ($(z1)+(0,-1)$) 
        {\textcolor{red}{\(X \not\!\perp\!\!\!\!\perp Y\)}\\ 
        \(X \not\!\perp\!\!\!\!\perp  Z\)\\ 
        \(Y \not\!\perp\!\!\!\!\perp  Z\)};

        \node[observed,label=above:{\(X\)}] (x2) at (2.5,0) {};
        \node[observed,label=above:{\(Y\)}] (y2) at (3.7,0) {};
        \node[observed,label=below:{\(Z\)}] (z2) at ($(x2)!0.5!(y2)+(0,-0.9)$) {};
        \draw[->,notouch] (y2) -- (z2);
        \draw[->,notouch] (z2) -- (x2);
        \node[align=center,scale=0.75] at ($(z2)+(0,-1)$) 
        {\textcolor{red}{\(X \not\!\perp\!\!\!\!\perp Y\)}\\ 
        \(X \not\!\perp\!\!\!\!\perp  Z\)\\ 
        \(Y \not\!\perp\!\!\!\!\perp  Z\)};

        \node[observed,label=above:{\(X\)}] (x3) at (5.0,0) {};
        \node[observed,label=above:{\(Y\)}] (y3) at (6.2,0) {};
        \node[observed,label=below:{\(Z\)}] (z3) at ($(x3)!0.5!(y3)+(0,-0.9)$) {};
        \draw[->,notouch] (z3) -- (x3);
        \draw[->,notouch] (z3) -- (y3);
        \node[align=center,scale=0.75] at ($(z3)+(0,-1)$) 
        {\textcolor{red}{\(X \not\!\perp\!\!\!\!\perp Y\)}\\ 
        \(X \not\!\perp\!\!\!\!\perp  Z\)\\ 
        \(Y \not\!\perp\!\!\!\!\perp  Z\)};

        \node[observed,label=above:{\(X\)}] (x4) at (7.5,0) {};
        \node[observed,label=above:{\(Y\)}] (y4) at (8.7,0) {};
        \node[observed,label=below:{\(Z\)}] (z4) at ($(x4)!0.5!(y4)+(0,-0.9)$) {};
        \draw[->,notouch] (x4) -- (z4);
        \draw[->,notouch] (y4) -- (z4);
        \node[align=center,scale=0.75] at ($(z4)+(0,-1)$) 
        {{\(X \perp\!\!\!\!\perp Y\)}\\ 
        \(X \not\!\perp\!\!\!\!\perp  Z\)\\ 
        \(Y \not\!\perp\!\!\!\!\perp  Z\)};
    \end{tikzpicture}
    \caption{V-structure example: only the rightmost causal structure satisfies all the dependences and independences between the variables $X, Y, Z$ ($X$ and $Y$ independent, $X$ and $Z$ dependent, and $Y$ and $Z$ dependent). The conditional independence statements in red are not consistent with the actual distribution described in the text.}
    \label{fig:vstructure}
\end{figure}
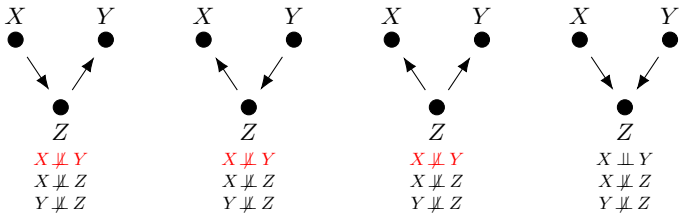

How can this principle help us identify a causal structure from observational data? We show a simple example that can be identified using this principle. Let us assume that we have three variables $X, Y$ and $Z$, for which we know that $X$ and $Y$ are independent, while the pairs $X$ and $X$  and  $Y$ and $Z$ are dependent. Following Reichenbach's principle and the causal faithfulness assumption, and assuming that there are no other variables involved, we can leverage these dependences and independences to identify that the only possible causal structure between $X, Y$ and $Z$ is the rightmost structure in Fig.~\ref{fig:vstructure}: because $X$ and $Y$ are independent, then $X$ and $Y$ do not cause each other, not even indirectly, nor are they both caused by another variable (in this case $Z$).

In this simple example, there is only one causal structure that fits the dependences and independences in the distribution given our assumptions. This is not the case in general, however. There are typically multiple causal structures that fit a distribution; these structures constitute a \emph{Markov Equivalence Class (MEC)}. If an edge between a variable $X$ and $Y$ are present in all of the structures in the MEC, we can say that the edge $X \to Y$ is identifiable, and $X$ has a direct causal effect on $Y$.

Causal discovery \citep{glymour2019review} provides a rich literature that allows us to learn causal relations in many complex settings, including under latent confounding. Extending Reichenbach's principle, \textit{constraint-based causal discovery methods}, e.g. PC/FCI \citep{spirtes2001causation}, leverage the assumption that conditional independence statements in the data correspond to certain patterns in the graphical structure that by the d-separation criterion \citep{Pearl2000-PEAC-2}. Through this correspondence, constraint-based methods identify the MEC by performing conditional independence tests on the data and using them as constraints on potential structures. Other methods instead focus on maximizing scores based on penalized likelihoods, e.g. \citep{chickering2002optimal,ramsey2017million}, or specific parametric assumptions \citep{hoyer2008nonlinear,shimizu2006linear}.

This is just a glimpse of causal discovery---it is an exciting and rapidly evolving field, and there are many interesting challenges that we will not have space to touch upon. A particularly relevant topic is causal discovery on time series, which provides more background knowledge information, \eg the temporal order of the causal variables can preclude certain types of graph. This also comes with additional challenges, \eg subsampling, time-varying confounding, and other types of non-stationarity. We refer to \citet{malinsky2018causal,glymour2019review,vowels2022d, zanga2022survey} for overviews of causal discovery.

\subsubsection{Challenges} 

Despite substantial recent progress in causal discovery in the last 30 years, many open challenges remain. We focus on three here.

\paragraph*{Reliability in Complex Settings.} 
Real-world scenarios often involve complex causal systems with many variables, some of which might not be observed (\iiee\ latent confounding), as well as 
selection bias, measurement noise or 
feedback loops.
Constraint-based methods provide some partial solutions to these issues as they can naturally model the presence of selection bias, latent confounding, or cyclic causal relations \citep{mooij2020constraint}. These methods can also integrate background knowledge, combine multiple datasets from different contexts \citep{mooij2020joint}, and combine data with different sets of overlapping variables \citep{triantafillou2015constraint}. Additionally, unlike other causal discovery methods, constraint-based methods shift any parametric assumptions to the specific conditional independence tests used, which can then also be nonparametric \citep{hsic}. 

At the same time, conditional independence testing---which is at the heart of constraint-based methods---is a notoriously difficult statistical problem, even for one test at a time \citep{Shah_2020,neykov2021minimax}. Constraint-based methods compound these issues by using multiple tests on the same data, for which error dependence is difficult to model even for simple structures \citep{cornia2014type}. Moreover, measurement error can introduce spurious dependence that these methods cannot detect; there is only limited work in these settings \citep{zhang2018causal}. 

An open challenge is which assumptions or models we can use to improve the reliability of the estimation of causal relations from data, \eg by improving the reliability of conditional independence tests, even in these complex settings. Another direction is improving the integration of potentially unreliable background knowledge from experts or AI systems such as Large Language Models; see Section \ref{sec:llm}.

\paragraph*{Scalability and Targeted Causal Discovery.} 
Scalability is a major challenge for all causal discovery methods. Most algorithms are NP-complete and their computational complexity scales exponentially with the number of variables. In many cases, however, we are only interested in the causal effects between a few \emph{target} variables and do not need to estimate all causal relations in a system.
We then only need to identify causal relations between the target variables and their (optimal) adjustment sets, which we can then use to estimate the causal effects.

Local causal discovery methods \citep[\eg][]{DBLP:journals/csda/WangZZG14,pmlr-v213-gupta23b} start to address this issue for a pair of target variables---a treatment and an outcome---using the parents of the treatment as an adjustment set. 
Other methods extend the types of adjustment sets, \eg by identifying groups of nodes that have certain properties in relation to this pair of targets \citep{pmlr-v244-maasch24a}.
\citet{watson2022causal} propose an algorithm to discover causal relations between multiple target variables under the restriction that the other variables are all non-descendants of the targets. Recently, \citet{pmlr-v258-schubert25a} remove this restriction by proposing an iterative version of the classic PC algorithm that sequentially identifies the non-ancestors of the targets by increasing the order of the conditional independence tests and removes them from consideration. This allows us to learn a subset of the graph that is relevant for statistically efficient adjustment sets, including ``optimal'' adjustment sets, \iiee sets that are optimal in terms of asymptotic variance \citep{henckel2022graphical}. On the other hand, this iterative approach still reconstructs a MEC on the whole set of ancestors of the targets, which might be more than necessary to identify only the optimal adjustment sets, so an open challenge is then how to learn only the minimal part of the causal graph that is relevant to estimate the causal effects among a set of target variables with optimal adjustment sets in a computationally efficient way.

\paragraph*{Post-selection inference in causal discovery.}
Another challenge in the integration of causal discovery and estimation is related to data ``double dipping'' \citep{gradu2022valid}, \iiee reusing the same data both for discovering the causal graph and for then estimating the causal effects, which invalidates the coverage guarantees of confidence intervals. Recently, \citet{chang2024post} proposed a first approach to solve this issue by resampling test statistics, which can also provide guarantees in case of incorrectly learned causal graphs. While exciting, this work 
mostly focuses on constraint-based methods and on settings without latent confounding. An open challenge is to extend this framework to more general settings and different causal discovery approaches.

\subsection{Aggregation and Synthesis of Causal Knowledge}

\subsubsection{Motivation}

A large body of causal inference research in the past few decades has focused on the design and analysis of a single study in isolation, often limited to binary treatments. 
However, a comprehensive understanding of causal relations ultimately relies on the integration, reconciliation, and synthesis of multiple study designs and data sources with multi-valued and multi-dimensional treatments.
A multi-faceted strategy that integrates multiple data sources and complementary study designs is therefore often necessary for causal inference in the real world.

\subsubsection{Background}

Consider, for example, a randomized experiment and an observational study.
Randomized experiments control biases by design and therefore yield effect estimates with high internal validity. But such studies are typically conducted on narrow samples of available study participants, which can limit the generalizability or external validity of their findings. 
Conversely, observational studies tend to encompass broader, potentially representative samples and accumulate larger datasets with richer covariate and temporal resolutions, yet generally possess lower internal validity because the treatment assignment is not controlled by the investigator. 

The causal inference literature has underscored the importance of integrating randomized and observational studies in enhancing the generalizability of experiments, improving their precision for targeted parameters, and reducing biases in observational studies \citep[\eg][]{brantner2023methods,colnet2024causal,dahabreh2024using}.
Doing so rests on strong identification assumptions, typically 
ignorability and overlap conditions on the treatment assignment and sample selection mechanisms; see \citet{hotz2005predicting, bareinboim2016causal} for a  discussion of identification for transporting and combining effects across different populations and settings.
Under these assumptions, estimation then proceeds via standard methods like treatment/selection weighting, outcome regression, and combinations thereof \citep[see, \eg][]{degtiar2023conditional}.
Finally, this literature is distinct but related to research on
inference from non-probability samples in sample surveys \citep[\eg][]{elliott2017inference}, integration of historical controls in clinical trials \citep[\eg][]{marion2023use}, and prediction under covariate shift in machine learning \citep[\eg][]{tibshirani2019conformal}.

\subsubsection{Challenges} \textcolor{white}{space}

\paragraph*{Estimating long-term effects.} 
A common problem faced by investigators is estimating the effect of a treatment on a long-term outcome, when the investigators only observe one or more short-term outcomes that are related to the primary outcome of interest.
Such short-term outcomes are often called ``surrogates'' for the long-term outcome \citep{prentice1989surrogate}.
\cite{athey2019surrogate} provide a framework for analyzing these issues, focusing on combining (1) an experimental study with the surrogate outcome only; and (2) a non-randomized study or descriptive data data set containing both the surrogate and primary outcome.
This is a relevant and recurring setting. 
For instance, in cancer research, the ultimate outcome of interest is often survival, but there are surrogates that can be measured earlier, such as tumor size or other indicators of disease progression. 
Similarly, in economics, the objective may be to measure long-term employment or income, but these are only available in administrative datasets.
In this setting, several challenges arise.
For example, \cite{athey2019surrogate} ask how to collectivize and systematize this scientific endeavor.
An open challenge is building a public library of surrogate indices for long-term outcomes that could expedite the analysis of future interventions.

\paragraph*{Constructing evidence factors.}
Drawing on \citet[\#265]{wittgenstein1958philosophical}  
regarding the critical distinction between new evidence and the same evidence repeated twice, \cite{rosenbaum2010evidence, rosenbaum2021replication} introduces a theory of evidence assembly in observational studies.
This theory aims to avoid the replication of biases that may have influenced earlier studies by isolating these biases, even though this process may expose them to new biases.
The aim is to ensure that the findings mutually reinforce each other.

Rosenbaum does this within the context of a single dataset, from which different study designs using various identification strategies are leveraged. 
Each of these studies produces a largely independent test for the null hypothesis of no treatment effect, which is affected by different forms of biases.
For instance, one can simultaneously exploit the assumption of unconfoundedness, an instrumental variable, and a dose-response relationship. 
This line of research offers promising opportunities, including the integration of alternative inferential approaches such as super-population sampling, the use of covariate adjustment techniques like weighting, and analytical tools from semiparametric efficiency theory.

\paragraph*{Using historical controls in clinical trials.} Another promising area that relates to the integration of diverse data sources is the use of historical controls. 
Randomization is the ideal method for evaluating treatments; however, in situations where it is unethical or impractical -- such as when an effective treatment already exists, the disease is rare, or the study population is particularly vulnerable -- investigators may have to rely on data from previous randomized experiments. 
These instances involve the use of historical controls, where previously collected data are used to construct comparison groups for evaluating new treatments.

This approach is gaining significant traction in pharmaceutical settings and raises several important methodological questions -— many of which intersect with broader issues in data fusion. 
For instance, how should we address covariate mismatch and non-compliance? 
How can we assess the validity of time-homogeneity assumptions that are typically in place? 
And how can we conduct efficient and unbiased estimation?
It is important to emphasize that the use of historical controls constitutes a non-experimental approach; therefore, results derived from such analyses should not be regarded as equivalent to those from randomized studies.

\paragraph*{Strengthening Causal Meta-Analyses.} Another major challenge arises when we have aggregated information from various studies and we need to synthesize it. 
Meta-analysis has a long history in statistical research \cite[\eg ][]{hedges2014statistical}, but it has developed largely independently from the field of causal inference, even though it is fundamentally concerned with causal questions. 
Notable exceptions include recent works by \citeauthor{dahabreh2020toward} [\citeyear{dahabreh2020toward}] emphasizing causal identification and semi-parametric estimation. 

Nevertheless, many challenges remain, such as synthesizing information from multiple experimental and non-experimental studies, each valid under different identification assumptions. 
How can we combine this evidence when some studies provide disaggregated data while others offer aggregated estimates? 
And how can we use this information to estimate action-relevant parameters, for instance, for a specific target population of patients?

\paragraph*{Analyzing complex systems.} Traditionally, causal inference studies have prioritized learning a single, well-defined causal estimand.
While this approach has produced valuable insights, a pressing challenge is how to causally describe an entire system, incorporating multiple dependencies and outcomes.
Consider, for example, a model of the whole economy such as the Phillips Moniac (Monetary National Income Analogue Computer)   \citep{bissell2007historical} or a mechanistic model of disease progression.
These models encompass interacting variables and outcomes.

Key questions in these domains concern the use of multiple data sources and quasi-experimental estimates to learn such models. 
The challenge lies in reconciling often disparate data sources and study designs and synthesizing this information to accurately describe the entire system. 
Of relevance here is \citeauthor{vanderweele2017outcome}'s [\citeyear{vanderweele2017outcome}] emphasis on ``outcome-wide'' epidemiologic studies rather than ``exposure-wide'' approaches, acknowledging that public health recommendations should be informed by a holistic understanding of how various factors collectively influence health outcomes.
By adopting such comprehensive approaches, we can better inform policy, ensuring that interventions are grounded in a robust understanding of the multitude of causal factors at play.

\paragraph*{Generalizing effect estimates when (some) causal relations are unknown.}
While we often assume that we know the causal relationships between the covariates, the treatment, and the outcome, in some settings they need to be learned from the data at hand.
This task is called \emph{causal discovery}, as we described in Section~\ref{subsec:causaldisco}. In the data fusion setting, all the challenges related to combining causal discovery and causal effect estimation in a statistically efficient way mentioned in Section~\ref{subsec:causaldisco} are compounded by additional challenges like multiple diverse populations with different distributions, or by datasets with covariate mismatch, {\it i.e.,} different variables measured in different settings.   

Previous work already provides some partial answers to this challenge. Causal data fusion \citep{bareinboim2016causal} is a principled framework to combine and transport causal effects across different populations. However, this approach requires a known causal graph and the knowledge of how distributions differ across the different populations. Although frameworks for causal discovery across multiple contexts \citep[\eg][]{mooij2020joint}, could be adapted to learn this knowledge from data, how to do this effectively is still an open question. Multiple methods have considered causal discovery in the challenging case of covariate mismatch or overlapping variables across datasets \citep{danks2008integrating,pmlr-v15-tillman11a,tillman2014learning,JMLR:v16:triantafillou15a,huang2020causal}. Another open question is how to combine causal discovery and causal effect estimation across populations. Recent work focused on observational data and experimental data \citep{pmlr-v206-triantafillou23a}, but how to do this effectively across different populations and with covariate mismatch remains an open question.

\subsection{Automation of the Causal Inference Pipeline}
\label{sec:automation}

\subsubsection{Motivation.}  
Manual derivations have long been the primary approach for obtaining identification results for causal effects, as well as for devising efficient estimators and valid confidence intervals for the corresponding target estimands. However, this approach is labor-intensive, often slow, and error-prone, as each new set of assumptions and target query requires a new derivation from scratch. Furthermore, this process ultimately constrains empirical analysts to mold their investigations around well-established pre-existing results---both in terms of assumptions and research questions---which may not always align with their domain knowledge.

The goal of automation in causal inference is to  provide scientists with methods that can flexibly adapt to their research needs. Automatic methods should handle various research questions, no matter how unconventional; leverage various types of domain knowledge, however imperfect; and, automatically handle model diagnostics as well as sensitivity analyses to key assumptions of the model.  Solving these tasks requires developing \emph{general purpose} tools that, given a set of causal estimands, assumptions and data, outputs (partial) identification results, along with finite sample estimates, confidence intervals, and statistical tests for any testable implications the model may entail.

\subsubsection{Background.} 

Automating causal inference tasks draws on various methods from symbolic and numeric computation, including techniques related to graph theory, matrix algebra, computer algebra, automated theorem proving, and optimization. Here we highlight a selected subset of these methods and applications.

One of the earliest automatic solutions to the problem of nonparametric identification originated in computer science with the development of the do-calculus and the ID algorithm, within the graphical models literature \citep{pearl1994probabilistic,tian2002studies}.  The ID algorithm provides a \emph{sound}, \emph{complete} and \emph{computationally efficient} solution for the problem of identifying interventional distributions from observational data, given assumptions encoded in a causal diagram \citep{huang2006pearl,shpitser2006identification}.\footnote{Soundness ensures that when the algorithm provides an answer, it is correct, while completeness guarantees that if the algorithm fails to do so, the query is provably not identifiable without additional assumptions or data. The ID algorithm runs in polynomial time with the number of variables.} This method belongs to the class of \emph{symbolic} approaches to identification, as it not only determines whether a causal effect is point-identified but also produces the corresponding identification formula. Since then, several algorithms have been developed, handling a rich variety of complex problems, such as selection bias, missing data, surrogate experiments, and generalizing results across experimental settings and domains (see, e.g. \citealp{bareinboim2016causal,mohan2021graphical}).  

Other symbolic approaches to automate identification problems leverage tools from computer algebra and automated theorem proving. For example, it can be shown that identification in linear structural equation models (SEMs) with Gaussian errors reduces to solving a system of polynomial equations. A sound and complete solution to this problem is  available using Gröbner bases and Buchberger’s algorithm \citep{garcia2010identifying}. Similarly, in nonparametric models with discrete variables, the problems of deriving testable implications   as well as (partial) identification of causal parameters can be solved using quantifier elimination algorithms \citep{geiger2013quantifier}. As another example, logic-based approaches based on SAT (Boolean Satisfiability) have been used to tackle a variety of causal inference problems, such as causal discovery, including in complex settings with latent confounding, cycles and selection bias, e.g. \citep{hyttinen2013}, and causal effect estimation with unknown graphs \citep{hyttinen2015}. However, in all these cases, the resulting algorithms are generally computationally inefficient. 

Beyond symbolic methods, numeric methods have also been used to solve identification problems in causal inference. One such approach is to recast the problem of (partial) identification as a \emph{constrained optimization problem}. For instance, in causal models with discrete observed variables, most queries, data, and assumptions can be expressed as polynomials of the probability mass function of discrete latent variables encoding all possible response types. The problem of (partial) identification is then equivalent to solving a \emph{polynomial program}. While solving polynomial programs is in general NP hard, methods exist that yield anytime valid bounds \citep{duarte2023automated}.\footnote{That is, if the algorithm runs long enough, it converges to sharp bounds. If it is stopped at any time, the bounds remain valid, but are conservative.} In contrast with the two previous examples,  here the final identification result is \emph{numeric} rather than symbolic.  In certain special cases, however, the problem can be simplified to a \emph{linear program}, for which efficient symbolic and numeric  algorithms are readily available \citep{balke1997bounds,sachs2023general}.

Once a parameter is (partially) identified, or testable implications are found, the next step in the causal inference pipeline is to draw inferences from finite data. The automation of this task requires the development of general-purpose  tools that can accept as input potentially arbitrary identifying estimands---be it in symbolic form, or even defined implicitly, such  as the solution to an optimization problem---and outputs (efficient) estimates as well as valid uncertainty quantification for those estimates.  The automation of estimation and inference has a long history in statistics; well known and widely used examples include M-estimation (coupled with numerical differentiation, automatic differentiation or the bootstrap) and probabilistic programming \citep{efron1994introduction,newey1994large,carpenter2017stan}. In some cases, such tools are already capable of handling estimands obtained from identification algorithms.  In many other cases, however, they either need to be adapted, or novel methods need to be developed.

\subsubsection{Challenges}
Despite significant advances in the automation of identification and estimation tasks, several challenges remain. We now outline some of the most pressing issues.

\paragraph*{Equality, inequality, and shape constraints.} Algorithmic approaches to identification have focused on leveraging exclusion restrictions and ignorability assumptions (\eg absence of arrows in a graph). However, other types of constraints often arise in causal inference. For instance, many popular identification strategies make use of shape constraints such as monotonicity (\eg\  no ``defiers''), equality (\eg\  ``parallel trends''), or continuity (\eg\  continuity of potential outcomes at a cut-off point). Additionally, sensitivity analyses impose (in)equality constraints on certain parameters. Preliminary progress has been made in incorporating monotonicity and equality constraints into graph-based approaches. See \citet{vanderweele2010signed,cinelli2019sensitivity,kumor2020efficient,zhang2021exploiting,maiticounterfactual} for examples with applications in (partial) identification and sensitivity analysis. In models with discrete variables, methods such as those discussed in the previous section can in theory handle such constraints, though in practice this has not yet been systematically studied, and issues of computational efficiency may arise.   How to systematically and efficiently leverage shape, equality, and inequality constraints remains an important topic of research, with many important applications.

\paragraph*{Scalability.} Scalability is another significant challenge for many automated procedures. For example, while methods from computer algebra, such as Gröbner bases, offer a complete solution for identification in linear systems, their complexity is doubly exponential in the number of variables.\footnote{As an illustration, it can take hours or days to solve a five variable linear model using Gröbner bases. Note, however, that a human may also take hours, days, if not months or years to solve the same model---if ever.}  Similar efficiency issues arise in current complete solutions for identification in discrete systems.  In many cases, it is possible to substantially speed up the performance of these algorithms by pre-simplifying equations or constraints. This is a pressing challenge to make such approaches more popular in practice. Developing algorithms that, while not complete, are efficient, also remains an important and active area of research for cases where a complete and efficient solution remains elusive or is outright known to be impossible.

\paragraph*{Hardness results.} Understanding the hardness of a task is important because it directly informs the computational resources required to solve it and guides algorithmic development. For example, if a task is known to be NP-hard, we may resort to approximation methods, heuristic approaches, or search for subsets of the problem for which an efficient algorithm may exist. 
Several identification problems still have unknown hardness.
For instance, when the causal graph is known, the identification of interventional queries can be solved in polynomial time in nonparametric models; however, at the time of writing, whether such an algorithm exists for linear models remains an open question.

\paragraph*{Knowledge representation.} Automating identification tasks requires a precise description of the query, the data and the assumptions the researcher is willing to defend.  How to systematically and efficiently represent the research question, the type of data collected, and various types of domain knowledge remains a significant challenge.  For example, causal graphical models provide a natural and intuitive way of systematically representing assumptions of  absence of direct effects (exclusion restrictions) or absence of confounding (ignorability restrictions); they have also been extended to encode missing data, measurement error, context specific independencies, or discrepancies across domains. On the other hand, many other types of domain knowledge remain under-explored, such as notions of variable importance, or shape constraints. Ideally, the profession should aim for a standard representation,  in a machine-readable file, of estimands, types of data, and assumptions, encoding all the meta-information needed to replicate the formal aspects of a study.

\paragraph*{Knowledge elicitation.} Even when the problem of representation is solved, knowledge elicitation remains a challenge. For example, most identification algorithms assume that researchers start from a fixed, known, set of assumptions. In reality, this is rarely the case. Researchers may possess various types of domain knowledge, some of which may aid in identification while much of it may be completely inconsequential. Additionally, distinguishing between useful and irrelevant knowledge can be difficult. Thus, knowledge elicitation can be viewed as a search problem. Automatic systems should assist empirical analysts in this search by offering methods that interactively help them elicit relevant knowledge reliably and efficiently.
Having a set of sufficient conditions for identification may also help with this process, as it can provide a useful starting point for analysis and data collection, which can then be refined (see Sections~\ref{sec:sensitivity}~and~\ref{sec:id-strategy}). 

\paragraph*{Automatic inference with machine learning.} 
The use of machine learning algorithms for causal effect estimation has gained significant attention. Constructing efficient estimators in these scenarios may require calculating a functional derivative known as the efficient influence function. Researchers have sought to automate these calculations through numerical \citep{frangakis2015deductive,luedtke2015discussion,carone2019toward,jordan2022empirical} and automatic \citep{luedtke2024simplifying} differentiation. Even when the analytical form of the efficient influence function is known, these methods depend on accurately estimating underlying nuisance parameters. To address this issue, loss functions have been developed to automatically estimate these nuisance parameters for regression functionals \citep{chernozhukov2022riesznet,chernozhukov2022automatic,van2025automatic}. Several open challenges remain in this area, including developing tailored loss functions that target general statistical functionals, automatically implementing estimators based on higher-order influence functions, or broadening the set of primitives to which automatic differentiation can be applied.

\paragraph*{Software.} Open-source software packages implementing many identification algorithms and estimation procedures are a fairly recent development. See, for example (non-exhaustive, and in alphabetical order): Ananke, autobounds, DoubleML, causaleffect, causaloptim, Causal Fusion, do-search, dagitty, EconML pcalg, tmle, among others \citep{textor2011dagitty,gruber2012tmle,kalisch2012causal,tikka2018identifying,karvanen2021search,syrgkanis2021causal,bach2022doubleml,jonzon2022accessible,lee2023ananke,duarte2023automated}. Notwithstanding this progress, software still remains an important bottleneck for the wide-spread adoption of algorithmic tools.

\paragraph*{Artificial intelligence.} While here we focused on the perspective of statisticians and data scientists who could benefit from using automated tools, the idea of automation of causal reasoning has its roots on the area of automated reasoning in artificial intelligence (AI).  Integrating causal reasoning tools with modern AI systems, such as large language models, is an exciting  direction of research that has not been fully explored, which we discuss in Section~\ref{sec:llm}.

\subsection{
Benchmarks, Evaluation, and Validation}

\subsubsection{Motivation}
A key driver of the rapid progress in machine learning and AI has been a shared focus on community benchmarks, such as ImageNet \citep{deng2009imagenet}, and the so-called common task framework \citep{donoho2023data}. In causal inference, by contrast, there have only been a handful of widely accepted community benchmarks. The few that exist, however, have had a major influence on the field---perhaps too much. We argue that developing new benchmarks and data challenges offers a promising opportunity for further advancement, with substantial demand from the causal inference community. Looking farther ahead, we also argue that the field needs to embrace more creative approaches for validating causal inference tools as applied in practice.

\subsubsection{Background.}
The primary hurdle to developing benchmarks is Holland's fundamental problem of causal inference. Unlike in machine learning, where there is typically a ``ground truth''---a correct image classification or future response variable---causality researchers can rarely have such access. This is because, with real-world data, we can observe only one of two or more counterfactual outcomes, making it difficult to precisely define a true causal effect. The field has therefore focused on several alternative approaches.

\paragraph*{Replicating experimental benchmarks.} In a hugely influential study, \citet{lalonde1986evaluating} sought to assess the empirical performance of regression methods then commonly used in economics. LaLonde began with a well-known experimental evaluation of a job training program, where the causal effect is estimated with high precision. 
He then replaced the experimental control group with individuals selected from surveys unrelated to the experiment. In the end, LaLonde found that regression using these non-experimental controls gave wildly different estimates than those from the original experiment itself, casting doubt on the suitability of these methods---and directly contributing to the rise in experimental evaluations of social policy. 
The primary challenge was the lack of overlap in covariate distributions in the two samples, see Figure 6, which displays histograms for the the log odds ratios, $\log ({\hat e}/({1 - \hat e}))$.
See \citet{dehejiawahba, imbens2024lalonde} for more details. 

A cottage industry within causal inference has conducted LaLonde-style replication exercises in settings ranging from voter mobilization \citep{arceneaux2006comparing} 
to online advertising \citep{gordon2019comparison}. More broadly, there is a large and growing literature on \emph{design replication} studies and \emph{within-study comparisons} \citep{wong2021design, chaplin2018internal}.

\begin{figure}[!th]
    \caption{Assessing Overlap in Lalonde-Dehejia-Wahba (LDW) Data}\label{fig:overlap}
    \centering
   
    \begin{minipage}[c]{.45\textwidth}
        \centering
        \captionsetup{justification=centering}
        \begin{subfigure}{0.75\linewidth}
            \includegraphics[width=\linewidth]{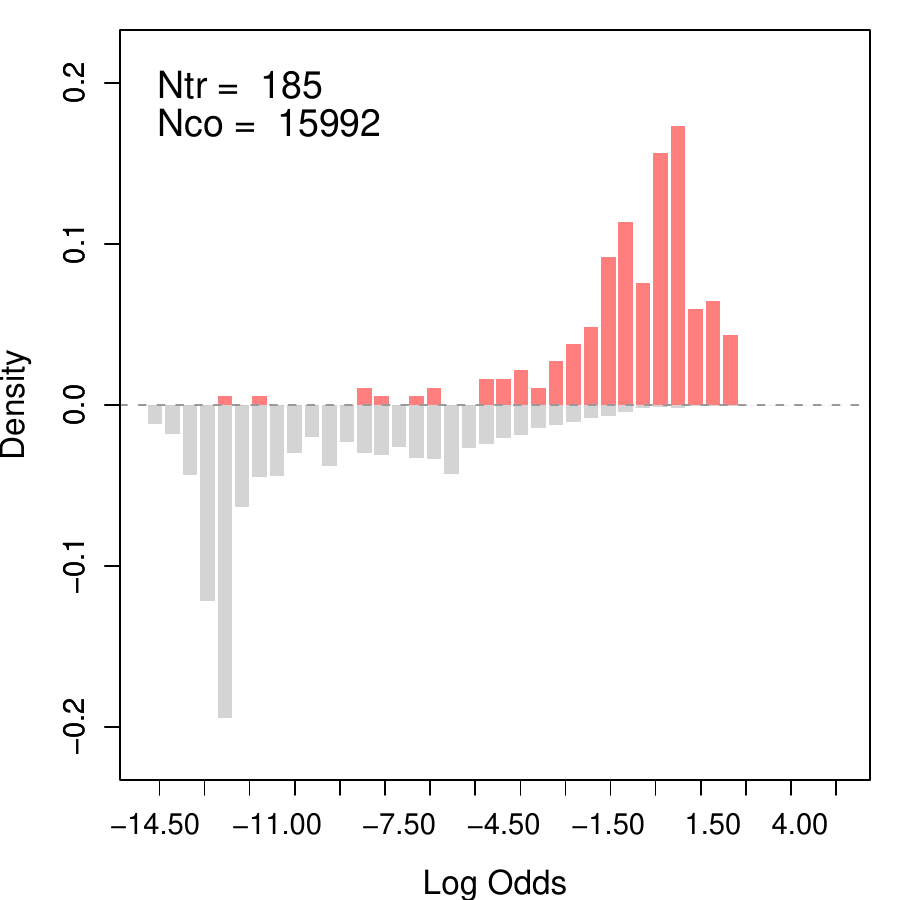}
            \caption{LDW-CPS}
        \end{subfigure}

    \end{minipage}%
    \\\raggedright
     {\footnotesize\textbf{\textit{Note:}} Histograms depict the log odds ratios, \iiee\ $\log({\hat{e}}/({1-\hat{e}}))$, using propensity score estimated through generalized random forest. The data are the observational CPS sample and the experimental from the National Supported Work Demonstration sample, both available in the Lalonde-Dehejia-Wahba data in \citet{dehejiawahba, imbens2024lalonde}.}
\end{figure}

\paragraph*{Synthetic and semi-synthetic benchmarks.} A common alternative is to construct benchmarks with simulated causal effects. \citet{kang2007demystifying} is a prominent, purely synthetic simulation benchmark that was developed to assess the performance of doubly robust methods. In another widely used benchmark, \citet{hill2011bayesian} constructed a semi-synthetic simulation, drawing the covariates, treatment, and control potential outcomes $Y_i(0)$ from the Infant Home Development Program (IHDP). More recently, the American Causal Inference Conference (ACIC) has hosted regular competitions where teams submit estimators with the goal of having the best predictions on a set of semi-synthetic datasets \citep{dorie2019automated}.

\paragraph*{Data challenges and common tasks.} Finally, researchers have organized around analyzing a common dataset, seeking to aggregate insights across different methods and research teams---without a (synthetic) ground truth. This has been more prominent within causal discovery for biology, such as 
the DREAM 5 causal discovery challenge, with the goal of estimating protein signaling networks in cells \citep{hill2016inferring}.

\subsubsection{Challenges.}  
While existing benchmarks are widely used---each with thousands of citations---relying too heavily on a single one of them, such as \citet{lalonde1986evaluating}, can distort incentives and lead the field to over-adjust to specific attributes of the study design. This highlights the need for new data challenges and benchmarks to further accelerate causality research.
In the short term, the causal inference community is actively working toward improved benchmarks and simulations that reflect the many and varied causal inference use cases \citep{curth2021really}. 
A longer-term goal is to validate the use of causal methods more broadly. 
Here the question is more fundamentally about science---\eg\ explanation, interpretation, and understanding mechanisms or causal structure---rather than prediction. As such, this is less amenable to ML-style benchmarks but remains critically important, especially after high-profile failures like those highlighted in the introduction.
Among other considerations, it is important to evaluate the methods not only in terms of their final estimates, but also to the extent to which they approximate features of a hypothetical randomized experiment and the robustness of their estimates.

\paragraph*{High-dimensional and complex data. } Existing causal inference evaluations have typically been quite low-dimensional---most re-analyses of \citet{lalonde1986evaluating} only have ten covariates, though there have been some high-dimensional variants \citep{farrell2015robust}. Such evaluations are out of step with the rapid rise of the application of machine learning methods, such as neural networks, to causal inference. Developing benchmarks with high-dimensional, and complex, multi-modal data remains an open challenge.

\paragraph*{Designed within-study comparisons.}
One promising generalization of LaLonde-style benchmarks is to explicitly incorporate within-study comparisons into a study design. 
For instance, \citet{shadish2008drpt} randomly assigned study participants to be in a randomized or nonrandomized trial, allowing for a direct assessments of non-experimental adjustment methods. Or \citet{saveski2017detecting} randomized whole groups into two different types of randomized designs to assess approaches for causal inference under interference.

\paragraph*{Tailored simulations.} A related issue is the gap between how causal inference methods are assessed when developed---typically in limited non-representative simulation studies and stylized data illustrations---and assessing the performance of that method in a specific application. One promising but underused strategy is to essentially design a simulation study tailored to specific applications, analogous to power calculations \citep{athey2021using}. Conducting such simulations can be prohibitive for domain scientists and can dramatically slow practical statistical analysis. Moreover, designing simulation studies specifically for causal inference presents a unique set of challenges \citep{evans2023parameterizing}. Thus, automating this process and lowering implementation barriers for domain scientists is an important goal \citep{nance2024causalsimulations}.

\paragraph*{Developing practical tools.} More broadly---and in the spirit of ``frictionless reproducibility'' in machine learning \citep{donoho2023data}---the causal inference community has only recently developed easily-modified repositories. For instance, an important setup was the \citet{dorie2019automated} R package for accessing semisynthetic data sets from the 2016 ACIC challenge; see also \citet{lin2019universal}. Further developing these practical tools will be critical for broader uptake.

\paragraph*{Ground truth validation in the biological and physical sciences.} Developing frameworks for validation necessarily requires creative solutions---and continued engagement with domain scientists. One direction is from biological applications, where researchers can sometimes pair data challenges and competitions with subsequent interventions to see which predictions were closest. In the context of gene knockout experiments, researchers might use a ``training set'' of interventions to predict promising interventions in a ``test set''---and then confirm those predictions with actual experimentation \citep[see][]{meinshausen2016methods}. In the context of physical systems, \citet{weichwald2022learning} challenged causal researchers to deploy methods on novel data from real-world control systems; and 
\citet{gamella2024causal} created computer-controlled laboratories that can assess and validate new causal methods. 
Finally, \citet{runge2019inferring} explored causal discovery methods in the context of climate data \citep[see also][]{brouillard2024landscape}.

\paragraph*{Improved success metrics.}
In many application domains, especially those in the medical and social sciences, the definition of ``success'' is open-ended. In addition to standard measures of statistical performance, other aspects, such as a strong study design and well-balanced covariates, can be crucial to the credibility and validity of a study; see, for example, \citet{zhao2019comment}. 
A close engagement with domain scientists is critical, as their specialized knowledge can guide the design and methods to address the most pertinent issues.

\paragraph*{Professional hurdles to building better benchmarks.} Even if the field agrees on the scientific goals, many professional and practical hurdles remain to building better benchmarks. Given the limited incentives for developing benchmarks, how can we secure funding to support their creation and maintenance? How can teaching efforts embed the development of benchmarks?
And how can domain experts be actively involved in this effort (\eg following the example of several ACIC data challenges)?

As with any approach, there will be fundamental limits to such benchmarking and validation. Understanding these limits will be important for making these approaches actionable in practice.

\subsection{New Identification Strategies}
\label{sec:id-strategy}

\subsubsection{Motivation.}

As we have discussed, going from association to causation requires a deep understanding of the mechanisms that generated the data. Typically, this understanding is encoded in terms of identification assumptions, which allow us to use observed measurements within the data to identify or ``see'' unobservable causal quantities that go beyond the data. In an ideal scenario, these assumptions are inherently valid by design, such as in the case of a perfectly conducted randomized experiment. However, as we depart from this ideal---especially in observational studies---these assumptions must be grounded in substantive considerations about the data, often referred to as an research design or identification strategy \citep{Angrist:Krueger:1999}.

There are now a fairly standard set of identification strategies widely used in the empirical social sciences, especially: adjusting or controlling for as many covariates as possible (see Sec 2.2); instrumental variables; regression discontinuity designs; and panel data strategies like difference-in-differences. 
This raises the question: besides these identification strategies, can we develop or uncover others?

\subsubsection{Background.}

\paragraph*{Synthetic controls.} A recent method that has gained much popularity is building synthetic controls \citep{abadie2003, Abadie2010}. This approach is often used in panel data settings to create a weighted average of control units that are balanced with respect to the pre-treatment outcomes of a focal treated unit. Since 2010, the number of applications of these methods in various forms has exploded in both academic settings and the private sector, especially in applications in which standard panel data methods are not appropriate.
There has been an active methodological literature exploring and extending this framework, including for uncertainty quantification and bias correction.

\paragraph*{Mendelian Randomization.} 

In the context of instrumental variables, Mendelian randomization (MR) has emerged as a promising strategy for uncovering causal effects \citep{sanderson2022mendelian}. Based on Mendel’s laws of inheritance, this strategy leverages the inherent randomness by which parents pass their genetic endowments to their offsprings. 
For example, MR has been used to explore the causal relationships between lifestyle factors such as alcohol consumption, diet patterns, and physical activity, and health outcomes such as cardiovascular disease.
Despite major methodological improvements, some open challenges remain.
Pleiotropic effects, where genetic variants influence multiple traits, can violate the exclusion restriction assumption, requiring the development of methods to detect and potentially adjust for these effects.
Weak genetic instruments can produce biased estimates and reduce statistical power, requiring robust methods that strengthen the relationship between the instrument and the exposure.
Furthermore, the selection and validation of genetic variants as instruments is crucial, requiring improved methods that integrate both biological and statistical perspectives.
Finally, comprehensive guidelines that delineate best practices for MR estimation are key to guide researchers in conducting robust analyses.

\paragraph*{Graphical models.} Graphical models may uncover novel identification opportunities beyond popular strategies, some of which may still be underexplored. As an example of a graphically inspired identification result, \citet{pearl1995causal} proposed the front-door criterion. This strategy has a similar structure to instrumental variables, in which three variables are linked in a causal path. Here the treatment of interest affects a mediator, the mediator affects the outcome, and, importantly, there is not direct causal path from the treatment to the outcome, other than through the mediator. The causal effect of the treatment on the outcome can then be identified through its mediating effects, even in the presence of unobserved treatment-outcome confounding, so long as the mediating paths are unconfounded. This causal model has not yet been widely explored in empirical social science, though see \cite{bellemare2024paper} for a recent application. More broadly, if a novel graph presents new identification results, there is a question of how frequently such situations occur in real-world settings. It is an interesting direction where mathematics offers insights into the real world, as opposed to the conventional approach of formalizing theories from empirical observations.

\subsubsection{Challenges.} 

\paragraph*{Instrumental variables.}
There are many open challenges in extending instrumental variable approaches.
Shift-share IVs, which have seen increased popularity in economics, exploit variation in the impact of external shocks across time and regions to identify causal effects \citep{borusyak2022quasi}. 
Separately, bunching techniques detect behavioral responses to policy changes through clustering in outcome distributions \citep{kleven2016bunching}. Both offer promising directions for further methodological research.

\paragraph*{Panel data.}

Another relevant direction is panel data methods, which exploit repeated measurements of stable units across time points \citep{arkhangelsky2024causal}. 
Historically, this literature has been largely distinct from the (mainly biostatistical) literature on time-varying treatments. Bridging these two literatures is a fruitful research direction. One recent example is for case-crossover designs, which provide a framework for examining transient exposures and acute events by comparing case (exposure) periods with control periods for the same individual \citep{shahn2023formal}. 
Exploring alternative identification strategies that involve forms of latent unconfoundedness offers interesting challenges for identification and inference \citep{athey2025identification}.

\paragraph*{Proximal methods.}

Additionally, the use of negative controls is gaining traction as a method for detecting and addressing confounding bias and model misspecification by leveraging variables or populations that are not expected to have causal effects as a benchmark \citep{tchetgen2014control}. These emerging methods collectively contribute to advancing proximal causal inference, a general framework that seeks to bridge different identification strategies \citep{tchetgen2024introduction}. 

\paragraph*{Combining strategies.}

As we have discussed, it is crucial to understand that the aforementioned strategies are not mutually exclusive.
Hence, an important question, is how to substantively combine them for more robust causal inference.

\subsection{Large Language Models and Causality}
\label{sec:llm}

\subsubsection{Motivation}
Large Language Models (LLMs) have burst onto the scene over the past several years, dramatically changing research across a wide range of domains in a short period. 
We are starting to see this impact in causal research as well. There is a growing literature at the interface of LLMs and causality, both on how to use LLMs to improve causal inference and on using ideas from causality for improving LLMs. 
Since this area is in flux, we offer a brief overview of some exciting topics here, recognizing that this literature is changing rapidly.

\subsubsection{Background and Challenges}

\paragraph*{Incorporating complex data into the causal inference pipeline.} Over the last decade, Natural Language Processing (NLP) methods have become increasingly common for leveraging text for causal inference, including as confounders, treatments, and outcomes \citep{veitch2020adapting,egami2022make,feder2022causal,imai2024causal_text}. For example, researchers might be interested in the effect of reading different messages on a downstream outcome, and then try to isolate key attributes that explain these effects \citep{gui2022causal,lin2024isolated}.
Building on these earlier ideas, the widespread availability of LLM tools has rapidly accelerated the use and scale of causal inference with text, as well as causal inference with other unstructured high-dimensional data, like images \citep{jerzak2023image, imai2025genai}.
Many open questions remain on how best to incorporate LLMs into the causal inference pipeline, including developing best practices for architecture and representation learning and appropriately quantifying uncertainty. Additional issues arise when LLMs and other generative AI systems are used to generate the interventions of interest---even defining the appropriate causal questions here is challenging.

\paragraph*{Synthetic experiments and Agentic Modelling.} Artificial or synthetic experiments are one explicitly causal area that has already seen preliminary work. Suppose one is interested in the effect of a particular intervention, say exposing individuals to a social media post about vaccines, on the likelihood they would subsequently take the vaccine. One could simply ask an LLM what the effect of such a post would be. More recent work however, has addressed such questions by creating a sample of artificial agents, and running an experiment on these agents, just like a regular experiment on real humans. See, for example,
\citet{hewitt2024predicting, de2025efficient, chen2025predicting, wang2025agenta, manning2024automated}. 
Such experiments seem to hold great promise in select settings given the speed with which one can conduct them and the avoidance of many (but not necessarily all) ethical issues.

At the same time, the performance of synthetic experiments in more realistic settings is an important open question; see \citet{gui2023challenge,anthis2025llm}. For example, in the parallel literature on synthetic audiences---using artificial agents to approximate survey respondents---there has been growing concern about out-of-sample performance \citep[\eg][]{li2025llm_persona_catch,von2025vox}. Going forward, it does seem likely that LLMs will be very helpful in augmenting randomized experiments. This can be in the form of performing initial runs using artificial agents, followed by targeted experiments using real individuals, or by replacing actual experiments entirely in some special cases. One challenge is to validate the results from artificial experiments, and to assess the uncertainty surrounding their results.  Another is to understand what features of such artificial experiments make them more or less reliable.

\paragraph*{LLMs and causal reasoning.} Whether LLMs have some version of causal reasoning  capabilities is an active research area \citep{kiciman2023causal,jin2023can_llms_infer}.
Examples include chain-of-thought reasoning \citep{wei2022chain} and hierarchical reasoning
\citep{wang2025hierarchical}. 
These questions are closely related to causal discovery and suggest many promising directions for incorporating LLMs to better understand complex causal systems. For example, an interesting question is whether current LLMs can discover causal relations from correlations \citep{jin2024can}, or if they can be used to provide common sense knowledge \eg as experts in the loop in combination with classic causal discovery tools \citep{ankan2025expertintheloop}  or as imperfect experts providing an initial causal ordering \citep{vashishtha2025causal}. Finally, there is also great interest in training \emph{transformers}, \iiee the backbone of many modern LLMs, as a foundation model to estimate causal effects \citep{ma2025foundation, balazadeh2025causalpfn}.

\paragraph*{AI co-pilots for causal inference.}  Echoing the use of AI tools to automate data analysis (see Section \ref{sec:automation}), there have been several recent efforts to use AI tools as ``co-pilots'' in aiding applied causal inference \citep{alaa2024large,verma2025causal_ai_scientist} and in study planning more broadly \citep{chang2024_nber_AI}. This direction shows promise as a means for democratizing applied causal inference research especially.

\paragraph*{Causality to improve LLMs.} The alternative direction has also shown great promise: using the ideas of causality and causal inference to improve AI development and deployment. 
For example, causal reasoning has proved important in developing more trustworthy and reliable AI deployments \citep{binkyte2025causality} and in better understanding preference learning methods used to train LLM \citep{lin2024optimizing,kobalczyk2025preference}.
Causal inference is also critical for evaluating and monitoring deployed AI systems \citep{liu2025bridging}. Another promising direction is the use of \emph{causal abstraction}, \iiee the theory describing the relations between different causal models at different granularities, as a way to interpret the internal mechanisms of the black-box LLMs, \eg  \citep{geiger2021,geiger2023,pmlr-v236-geiger24a}.

\paragraph*{Open questions.}
LLMs have burst onto the scene in a very short time, and have already had major impacts on many areas of study. The impact on causality and causal inference is still limited at the moment, but likely to be profound. Exactly where this will be is a difficult question to answer, and we are excited to see what develops.

\section{Concluding thoughts}
\label{section:thoughts}

For centuries, the quest to establish causation has been central to scientists in their efforts to uncover the laws that govern the natural world.
In today's society, a similar aspiration drives policy makers: to find which interventions work to build a better social world.
We seek to understand causation, uncover natural laws, and learn which interventions work through data from the real world---a process known as causal inference. 
This is a perennial quest that will never end.

\subsection{Many additional directions}
In this paper, we explore this issue from the perspective of statistics and related fields, yet there is much to gain integrating diverse perspectives and fostering cross-disciplinary work. 
While we provide an overview and discuss some of the open challenges in causal inference from this perspective, this is merely a partial map of a much broader landscape.
There are many others that we have not addressed. 

\paragraph*{Time and dynamics.} One such challenge is incorporating more complex temporal dimensions. While we covered various principles and problems that generally apply to causal studies with increased temporal measurements, this dimension adds a specific complexity. Despite the 40 years of progress since \citeauthor{robins1986new}'s [\citeyear{robins1986new}] foundational paper, integrating time-dependence---encompassing treatments, covariates, outcomes, and their feedback---into routine causal analyses remains a persistent challenge.

\paragraph*{Bayesian causal inference.}
Another challenge involves using Bayesian methods for causal inference.
Although Bayesian approaches have been proposed and adopted, especially in the literature on heterogeneous treatment effects and causal machine learning \citep[\eg][]{hill2011bayesian, hahn2020bayesian}, the frequentist paradigm has traditionally dominated the field.
Useful perspectives on Bayesian methods for causal inference are provided in, among others, \citet{li2023bayesian} and \citet{daniels2023bayesian}.

\paragraph*{Designing observational studies and target trial emulation.}
Furthermore, a core idea in causal inference is to cast causal questions in terms of concrete interventions and conceptualize them in the context of hypothetical randomized experiments.
This notion has a long history, dating back at least to the work of \cite{dorn1953} and \cite{cochran1965}.
In this tradition, one widely adopted framework in the health sciences is that of target trial emulation \citep{hernan2016using}.
A key challenge is how to effectively apply this approach in other scientific domains and policy contexts.

\paragraph*{Causes of effects.} Finally, while we have focused on the question of the effect of causes, the reverse question of the causes of effects is also important \citep{dawid2022effects}.
This type of questions emerges often in legal contexts, for instance, in determining whether a certain drug caused a health problem.

\medskip
All these challenges and possible research directions intersect and compound each other, opening up new opportunities to advance causal inference.

\subsection{Cross-cutting issues}
Our main discussion raises several cross-cutting issues that we want to emphasize again here.

    \paragraph*{Bridging the gap between theory and practice.} Every challenge we discuss above faces a familiar disconnect between applications and theory. There is an urgent need to make state-of-the-art methods accessible and applicable to address substantive research problems in practice and to facilitate their adoption.
    This requires deep engagement with substantive experts, the development of better software and practical diagnostics, and better guidance and outreach.  
    In the other direction, such engagement will hopefully lead to new and exciting methodological challenges, following a long tradition of advances in causal inference.

    \paragraph*{Gains from incorporating machine learning and computational tools.} Over the last two decades, the widespread availability of improved computation and machine learning methods has led to fundamental changes in the practice of causal inference and launched the increasingly influential subfield of ``causal machine learning.'' Today, powerful LLM and AI tools offer similar promise, and we anticipate exciting uses for these new technologies in causality research.

    \paragraph*{Importance of building an open and inclusive research community.} As we outline above, causality research has historically been highly fractured, with often distinct and isolated research communities. 
    Over the last decade especially, there has been substantial progress in creating a ``big tent'' causal community. In addition to making the field more welcoming and inclusive, this has also led to important research advances and more rapid adoption of novel ideas. Furthering such efforts is critical for the continued growth and evolution of causality research.

\subsection{Asking good questions} 
In the end, the central challenge in causality research is to continue to ask good questions. We have posed many questions for the causal community here. But as the field continues to grow, what questions are we missing or failing to ask?

\clearpage
\begin{acks}[Acknowledgments]
The authors would like to thank the anonymous referees, an Associate
Editor and the Editor for their constructive comments that improved the
quality of this paper.
\end{acks}

\begin{funding}
Carlos Cinelli was supported in part by the Royalty Research Fund at the University of Washington, and by the National Science Foundation Grant No. MMS-2417955.


Guido Imbens was partially supported by the Office of Naval Research under grant numbers N00014-17-1-2131 and N00014-19-1-2468 and a gift from Amazon.

\end{funding}


\bibliographystyle{abbrvnat} 
\bibliography{refs}       

\vfill\eject

\end{document}